# On Metaverse Application Dependability Analysis

Yingfan Zong, Jing Bai, Xiaolin Chang, Fumio Machida, Yingsi Zhao

**Abstract**—Metaverse as-a-Service (MaaS) enables *Meta*verse tenants to execute their *APP*lications (MetaAPP) by allocating Metaverse resources in the form of Metaverse service functions (MSF). Usually, each MSF is deployed in a virtual machine (VM) for better resiliency and security. However, these MSFs along with VMs and virtual machine monitors (VMM) running them may encounter software aging after prolonged continuous operation. Then, there is a decrease in MetaAPP dependability, namely, the dependability of the MSF chain (MSFC), consisting of MSFs allocated to MetaAPP. This paper aims to investigate the impact of both software aging and rejuvenation techniques on MetaAPP dependability in the scenarios, where both active components (MSF, VM and VMM) and their backup components are subject to software aging. We develop a hierarchical model to capture behaviors of aging, failure, and recovery by applying Semi-Markov process and reliability block diagram. Numerical analysis and simulation experiments are conducted to evaluate the approximation accuracy of the proposed model and dependability metrics. We then identify the key parameters for improving the MetaAPP/MSFC dependability through sensitivity analysis. The investigation is also made about the influence of various parameters on MetaAPP/MSFC dependability.

## 1 Introduction

With recent technological advancements (e.g., extended reality, 5G/6G networks, and edge intelligence) and the substantial endeavors of major corporations like Facebook [1] and Microsoft [2], Metaverse is being garnering increasing attention from both academia and industry over the past years. Metaverse as-a-Service (MaaS) [3] sprouts up due to the fact of dynamic and huge resources required in an *Meta*verse *APP*lication (MetaAPP), e.g., education, healthcare, and manufacturing [4]. MaaS not only allows Metaverse users/tenants to pay on demand, but also improves both Metavserse service dependability (availability and reliability) and Metaverse resource utilization.

It is true that both Metaverse resource demands from Metaverse tenants and available resources of MaaS provider are dynamic and uncertain. Decomposing an MetaAPP and then executing it in the form of Metaverse service functions (MSF) can alleviate the impact of these dynamicity and uncertainty on MetaAPP dependability [5], namely, the dependability of the MSF chain (MSFC), consisting of MSFs allocated to MetaAPP. MSFC and MetaAPP are used interchangeablly in this paper. Fig. 1 illustrates three MetaAPPs/MSFCs in edge-cloud-supported MaaS. MaaS provider applies Edge-Cloud resources to construct an MSFC for each MetaAPP from end users.

Each MSF is usually deployed in a virtual machine (VM) for better resiliency and security [6]. However, these MSFs along with VMs and virtual machine monitors (VMMs) executing them cannot avoid software aging after prolonged continuous execution. Then MSFC dependability (including availability and reliability) decreases and even the MSFC crashes [7][8]. Software aging can be handled by rejuvenation techniques like MSF failover, VM failover and VM migration. But they require the support of backup components, which also have the problem of software aging.

This paper explores analytical modeling techniques to quantitatively study the impact of both software aging and rejuvenation techniques on VM-based MSFC dependability in the scenarios, where both active components (MSF, VM and VMM) and their corresponding backup components are subject to software aging. We assume that the MSFC are established. How to decompose MetaAPP to set up an MSFC is beyond the scope of this paper.

To the best of our knowledge, it is the first time to quantitatively investigate VM-based MSFC dependability. There existed analytical models [9]-[22] for evaluating service function chain (SFC), which is similar to MSFC. However, the models in [9]-[22] were not suitable for analyzing time-dependent interactions between backup and active component behaviors, and time-dependent interactions between MSF, VM and VMM behaviors. And the authors in [9]-[22] ignored the behaviors of backup MSFs, backup VMs and backup VMMs. There are at least the following two challenges to be addressed in employing analytical model-based approaches to assess MSFC dependability.

- The time-dependent interactions between active component behaviors and backup component behaviors in the VM-based MSFC are more complex than that in the container-based MSFC. That is, there are backup MSFs, backup VMs and backup VMMs in the VM-based MSFC, and their behaviors interact with the behaviors of active MSFs, active VMs and active VMMs. In addition, these interactions are time-dependent. Therefore, capturing the behaviors of all components and the time-dependent interactions between various behaviors is a challenge.

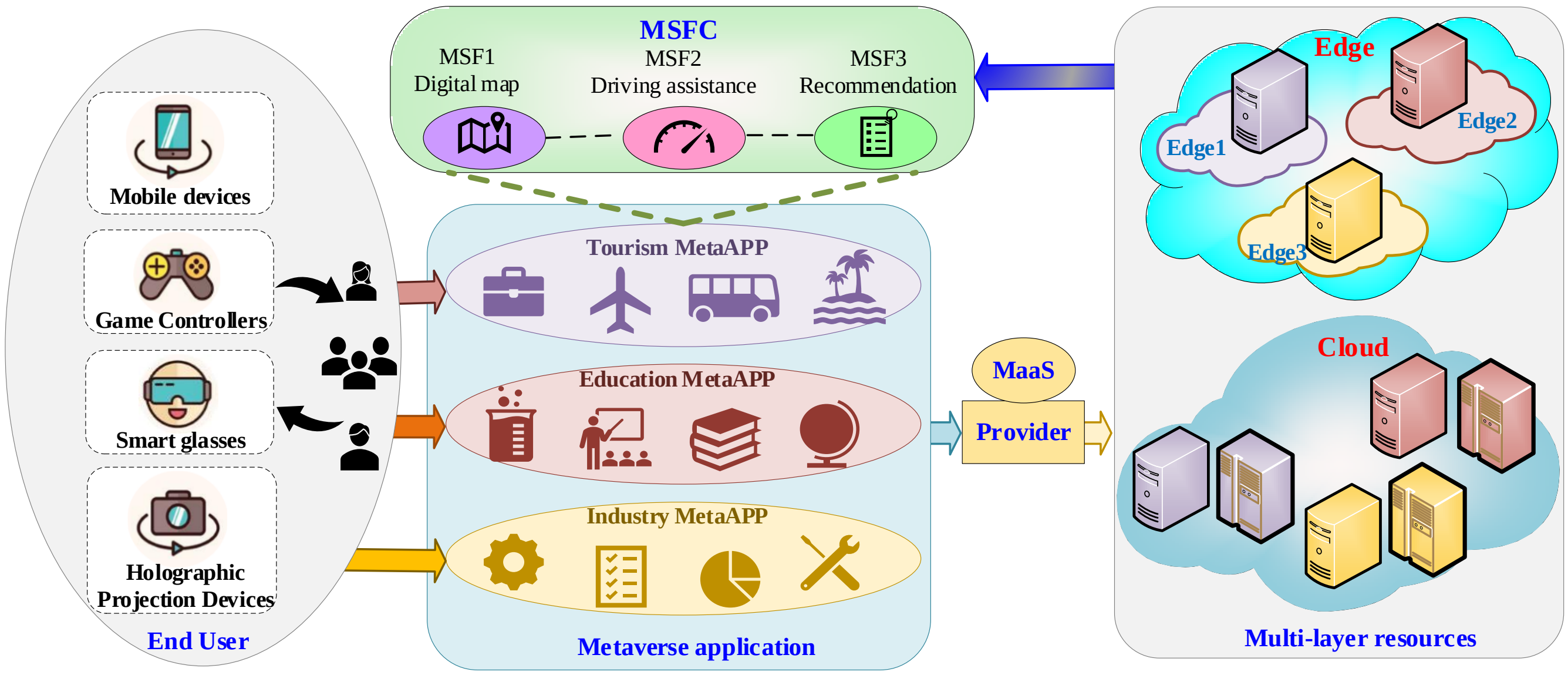

Fig. 1 MetaAPP over Edge-Cloud-supported MaaS

- Rejuvenation-triggered intervals (RTIs) can affect the effectiveness of rejuvenation techniques. However, various time-dependent interactions in the VM-based MSFC are more complex than that in the container-based MSFC, making it difficult to collaboratively determine RTIs of MSF failover, VM failover and VM migration. Therefore, how to determine the optimal combination of RTIs of MSF failover, VM failover and VM migration and its corresponding optimal dependability is another challenge.

In order to tackle the above challenges, we explore a hierarchical model to quantitatively analyze the availability and reliability of MSFC system, incorporating proactive rejuvenation techniques to alleviate components' aging issues. The principal contributions of this paper can be summarized in the following manner:

- We develop a hierarchical model for an MSFC with $n$ MSFs. The model includes $n$ semi-Markov process (SMP) sub-models and a reliability block diagram (RBD) sub-model. Each SMP sub-model captures the behaviors of components in a primary Metaverse host and its backup Metaverse host (See Fig. 2). The RBD sub-model describes the composition of $n$ SMP sub-models. We focus on the transition from the occurrence of an abnormal event (software aging or failure) to recovery through rejuvenation techniques. In our SMP sub-model, failure and recovery event-occurring times follow non-exponential distribution.
- We assess VM-based MSFC dependability in terms of steady-state availability and reliability measured by mean time to MSFC failure (MTTF). In particular, under the consideration of RTIs, we derive the closed-formed formulas for analyzing the steady-state availability and reliability of an MSFC comprising arbitrary number of MSFs.

We verify the validity of the proposed model and formulas through creating a simulator and conduct extensive numerical experiments. The experimental results indicate:

1) The cumulative distribution function (CDF) type of failure time, active MSF failure time and host fix time are important factors for improving MSFC dependability.

2) As the number of serial MSFs increases, the MSFC dependability decreases. Conversely, as the number of parallel MSFs increases, the MSFC availability increases while the MSFC reliability decreases.

3) There exists the optimal RTI combination that can maximize the MSFC dependability.

4) There is a significant difference between the numerical results under the model with considering the backup components' behaviors and those under the model without considering the backup components' behaviors.

The rest of the paper is set up as follows. Section 2 introduces related work. Section 3 introduces the studied MSFC system, the proposed hierarchical model and the formulas of calculating availability and MTTF. Section 4 presents the experiment results. Section 5 concludes the paper.

## 2 RELATED WORK

Our investigation of public literatures finds no analytical model-based study on MSFC dependability. Thus, this section first introduces the existing researches of Metaverse, and then analytical model-based approaches for SFC are discussed.

### 2.1 Metaverse

Metaverse is a living space and cyberspace that realizes the virtualization and digitization of the real world [23]. The current Metaverse research focuses on the following areas: constructing blockchain-based economic ecosystems [24], offering immersive experiences through interactive technologies [25], generating mirror images of the real world based on digital twins [26], accomplishing data computation, storage, processing and sharing through cloud computing [27], and achieving interconnected intelligence through AI [28] and IoT technologies [29]. However, we are the first to model the MSFC system behaviors for the dependability assessment. Our work complements the existing Metaverse works for better MaaS service provision.

TABLE 1 COMPARISON OF EXISTING MODELS

| Reference | SF | VM | VMM | SFC system characteristic | | | | Model | Distribution | | Metric | | Simulation | Sensitivity Analysis |
|---|---|---|---|---|---|---|---|---|---|---|---|---|---|---|
| | | | | Aging Behavior | Behaviors of Backup SFs, VMs and VMMs | Time Dependence | RTI | | Exponential | General | Availability | Reliability | | |
| [9] Zhang, 2019, [10] Wang, 2021 | √ | × | × | × | × | × | × | RBD | × | × | √ | × | × | × |
| [11] Wang, 2023 | √ | × | × | × | × | × | × | × | × | × | √ | × | × | × |
| [12] Rui, 2021 | √ | × | × | × | × | × | × | CTMC | √ | × | × | √ | × | × |
| [13] Simone, 2023 | √ | √ | × | × | × | × | × | CTMC | √ | × | √ | × | × | × |
| [14] Tola, 2021 | √ | √ | × | √ | × | × | × | CTMC | √ | × | √ | × | × | √ |
| [15] Bai, 2023 | √ | × | × | √ | × | × | × | SMP | √ | √ | √ | √ | √ | √ |
| [16] Bai, 2021 | √ | × | × | √ | × | × | × | SMP | √ | √ | √ | √ | √ | × |
| [17] Bai, 2022 | √ | × | × | √ | × | × | √ | SMP | √ | √ | √ | √ | √ | √ |
| [18] Bai,2023 | √ | × | √ | √ | × | × | × | SMP | √ | √ | √ | √ | √ | √ |
| [19] Bai,2023, [20] Bai,2023 | √ | × | √ | √ | × | × | √ | SMP | √ | √ | √ | √ | √ | √ |
| [21] Mauro,2021 | √ | √ | √ | × | × | × | × | FT+CTMC | √ | × | √ | × | × | √ |
| [22] Pathirana,2023 | √ | √ | √ | × | × | × | × | FT+CTMC | √ | × | √ | × | × | × |
| Ours | √ | √ | √ | √ | √ | √ | √ | √ | √ | √ | √ | √ | √ | √ |

[a] 'Aging Behavior' indicates whether the impact of software aging is considered.
[b] 'Behaviors of Backup SFs, VMs and VMMs' indicates whether the behaviors of backup SFs, VMs and VMMs are analyzed.
[c] 'Time Dependence' indicates whether there are time-dependent interactions between SF, VM, and VMM behaviors.
[d] 'RTI' indicates whether rejuvenation techniques are triggered immediately or after a certain interval time when software aging occurs.
[e] 'Model' indicates which the state-space models are constructed.
[f] 'Distribution' indicates whether the event-occurring times follow exponential distribution or general distribution.
[g] 'Metric' indicates whether availability and reliability are evaluated.
[h] 'Simulation' indicates whether simulation is performed for the model verification.
[i] 'Sensitivity Analysis' indicates whether the sensitivity analysis is carried out.

## 2.2 Backup Strategies for SFC Dependability Improvement

Zhang et al. [9] considered the dedicated backup and shared backup strategies and investigated the resource-aware backup allocation problem with the goal of minimizing backup resource consumption while satisfying availability. Wang et al. [10] suggested utilizing a backup SFC to safeguard the active SFC, thereby improving the availability of parallelized SFCs, and assessed the availability of the SFC. Wang et al. [11] took a comprehensive approach, considering from both the end users and edge system to backup SFC, in order to provide services with the lowest latency. Our work can evaluate the availability and reliability of MSFC, which is complementary to the aforementioned studies, so as to help service providers provide better services.

## 2.3 Model-based Approaches on SFC Dependability

The main goal of our work is to evaluate the availability and reliability of MSFC system based on analytical models. We discuss the existing modeling approaches in the following three categories: state-space models, non-state-space models, and multi-level models.

### 2.3.1 Non-state-space Model

Non-state-space models mainly include: RBD, reliability graph and fault tree (FT). Zhang et al. [9] studied SFC availability based on RBD. The authors in [10] used RBD to model different SFC configurations and analyzed the parallelized SFC availability. The authors in [9] and [10] assumed that the behaviors of individual components in a host are independent. However, in practical systems, there may exist dynamic interactions between the abnormal behaviors and the recovery behaviors of each component. The model introduced in this paper has the capability to encompass the time-dependencies among these behaviors.

### 2.3.2 State-space Model

Rui et al. [12] investigated the SFC reliability based on Petri net model and further designed the VNF migration strategy. Simone et al. [13] proposed the continuous-time Markov chain (CTMC) model for analyzing the multi-tenant SFC availability. Tola et al. [14] proposed the stochastic activity network (SAN) models to describe different network function virtualization management and orchestration (NFV-MANO), and assessed the impact of software rejuvenation on the NFV-MANO availability. Studies [12]-[14] assumed that all event-occurring times followed the exponential distribution. For the past few years, our research teams have developed various state-space models based on SMP. Studies [15] and [16] respectively investigated the dependability of serial SFC and hybrid serial-parallel SFC. Subsequently, the study [17] considered the impact of RTIs on the container-based SFC dependability. In addition, studies [18] and [19] extended the development of SMP models to capture the behaviors of multiple SFs and operating systems (OSes) in a serial SFC system and a hybrid serial-parallel SFC system, respectively. Furthermore, they [20] examined the influence of backup component behaviors on the container-based SFC dependability. The above studies considered the effect of software aging and assumed that the failure and recovery event-occurring times followed general distribution. But they cannot capture the time-dependent interactions between MSF, VM and VMM behaviors in the MSFC system.

#### 2.3.3 Multi-level Model

Besides the non-state space and state-space modeling approaches, there are multi-level modeling approaches to evaluate the dependability of virtualized systems. Mauro et al. [21] evaluated the SFC availability and performance by aggregating stochastic reward networks (SRN) and RBD. Pathirana et al. [22] calculated the availability for 5G-MEC systems based on FT and SAN. Compared with these studies, the model constructed in this paper can describe the time-dependent interactions between active component behaviors and backup component behaviors under the condition of non-exponential failure and recovery event-occurring times.

TABLE 1 summarizes the comparison of the existing works about SFC dependability analysis by analytical model-based approaches.

## 3 System Description And Hierarchical Model

This section first introduces the serial and parallel MSFC system architectures. Then the hierarchical model and metric formulas are presented. TABLE 2 and TABLE 3 give the definition of variables used in the model.

### 3.1 System Description

The MSFC system investigated in the paper consists of a Control Plane, several Primary and Backup Metaverse hosts. Control Plane is responsible for creating, monitoring and maintaining the MSFC system. Each host only runs one VMM. In a Primary Metaverse host, the VMM can deploy an active VM and a backup VM, which execute an active MSF and a backup MSF, respectively. Backup MSFs and backup VMs are deployed to support failover technique. Backup VMMs in Backup Metaverse hosts are deployed to support VM migration technique.

Fig. 2 MSFC illustrates an MSFC system with four MSFs and also shows two examples of MSFC system architecture. One is a serial MSFC system architecture consisting of four serial MSFs. Another example is a parallel MSFC system architecture consisting of two serial MSFs and two parallel MSFs, where serial MSF1 runs on Primary Metaverse host1, parallel MSF2 runs on Primary Metaverse host2, parallel MSF3 runs on Primary Metaverse host3, and serial MSF4 runs on Primary Metaverse host4. The OS in the figure represents the operating system.

In an MSFC system, we consider that the active MSFs, backup MSFs, active VMs, backup VMs, active VMMs and backup VMMs can suffer from software aging. In addition, when active MSF, active VM or active VMM aging is detected, rejuvenation technique is not triggered immediately but wait for a while before triggering.

At the beginning, all MSFs, VMs and VMMs work properly. After a period of time, MSFs, VMs and VMMs start to suffer from software aging. When one of the active components is detected to be aging, the MSFC system immediately checks the state of the corresponding backup component for that active component. There are three states of a backup component:

- Backup component being healthy. This backup component can take over the work of the active component. The rejuvenation technique (MSF failover/VM failover/VM migration) is triggered after waiting for a period of time.
- Backup component suffering from software aging. This backup component will be restarted immediately to be ready to support rejuvenation technique. After restarting the backup component and waiting for a while, rejuvenation technique will be triggered.

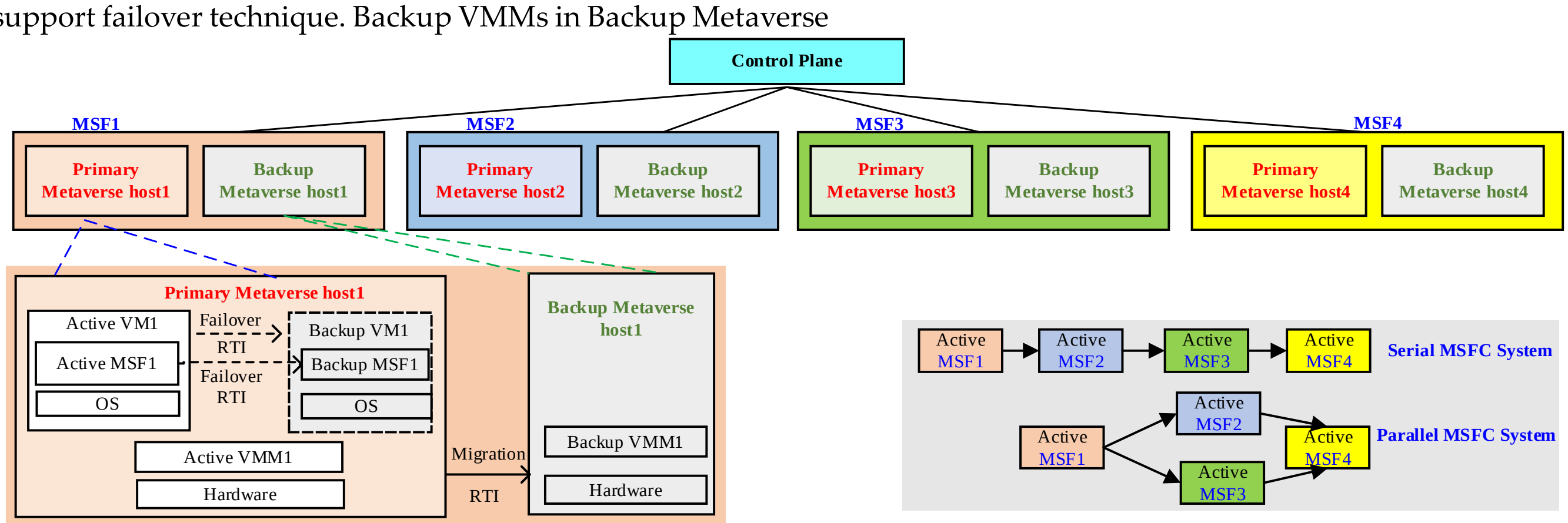

Fig. 2 MSFC system

TABLE 2 Definition of Variables Used in the Model and Experiments

| Symbol | Definition | Default Value |
|---|---|---|
| $n$ | The number of MSFs in the MSFC system. | 4 |
| $m$ | The number of MSFs in the serial part of the parallel MSFC system we study. | 2 |
| $c_{S1}$ / $c_{V1}$ / $c_{M1}$ | The probability of backup MSF/backup VM/backup VMM being at H state after active MSF/active VM/active VMM being detected to suffer from software aging. | 0-1 |
| $c_{S2}$ / $c_{V2}$ / $c_{M2}$ | The probability of backup MSF/backup VM/backup VMM being at D state after active MSF/active VM/active VMM being detected to suffer from software aging. | 0-1 |
| $c_{S3}$ / $c_{V3}$ / $c_{M3}$ | The probability of backup MSF/backup VM/backup VMM being at F state after active MSF/active VM/active VMM being detected to suffer from software aging. | 0-1 |

TABLE 3 DEFINITION OF VARIABLES USED IN THE MODEL AND EXPERIMENTS

| Symbol | Definition | Dis. | Default Value |
|---|---|---|---|
| $T_i^{aas}$ | A variable with CDF $F_i^{aas}(t)$ denoting the time of the $i^{th}$ active MSF from H to D state. (Active MSF aging time) | EXP | 18-30 months |
| $T_i^{aav}$ | A variable with CDF $F_i^{aav}(t)$ denoting the time of the $i^{th}$ active VM from H to D state. (Active VM aging time) | EXP | 24-36 months |
| $T_i^{aam}$ | A variable with CDF $F_i^{aam}(t)$ denoting the time of the $i^{th}$ active VMM from H to D state. (Active VMM aging time) | EXP | 30-42 months |
| $T_i^{abs}$ | A variable with CDF $F_i^{abs}(t)$ denoting the time of the $i^{th}$ backup MSF from H/BR/BC to D state. (Backup MSF aging time) | EXP | 18-30 months |
| $T_i^{abv}$ | A variable with CDF $F_i^{abv}(t)$ denoting the time of the $i^{th}$ backup VM from H/BR/BC state to D state. (Backup VM aging time) | EXP | 24-36 months |
| $T_i^{abm}$ | A variable with CDF $F_i^{abm}(t)$ denoting the time of the $i^{th}$ backup VMM from H/BR/BC state to D state. (Backup VMM aging time) | EXP | 30-42 months |
| $T_i^{fsa}, T_i^{fsr}, T_i^{fsc}, T_i^{fsd}$ | The variables with CDFs $F_i^{fsa}(t)$, $F_i^{fsr}(t)$, $F_i^{fsc}(t)$ and $F_i^{fsd}(t)$ denoting the time of the $i^{th}$ active MSF from D to F state under the corresponding backup MSF at A, BR, BC and D state, respectively. (Active MSF failure time) | G | 22-26 months |
| $T_i^{fsl}$ | A variable with CDF $F_i^{fsl}(t)$ denoting the time of the $i^{th}$ active MSF from L to F state. (Active MSF failure time) | G | 22-26 months |
| $T_i^{fva}, T_i^{fvr}, T_i^{fvc}, T_i^{fvd}$ | The variables with CDFs $F_i^{fva}(t)$, $F_i^{fvr}(t)$, $F_i^{fvc}(t)$ and $F_i^{fvd}(t)$ denoting the time of the $i^{th}$ active VM from D to F state under the corresponding backup VM at A, BR, BC and D state, respectively. (Active VM failure time) | G | 34-38 months |
| $T_i^{fvl}$ | A variable with CDF $F_i^{fvl}(t)$ denoting the time of the $i^{th}$ active VM from L to F state. (Active VM failure time) | G | 34-38 months |
| $T_i^{fma}, T_i^{fmr}, T_i^{fmc}, T_i^{fmd}$ | The variables with CDFs $F_i^{fma}(t)$, $F_i^{fmr}(t)$, $F_i^{fmc}(t)$ and $F_i^{fmd}(t)$ denoting the time of the $i^{th}$ active VMM from D to F state under the corresponding backup VMM at A, BR, BC and D state, respectively. (Active VMM failure time) | G | 46-50 months |
| $T_i^{fmm}$ | A variable with CDF $F_i^{fmm}(t)$ denoting the time of the $i^{th}$ active VMM from M state to F state. (Active VMM failure time) | G | 46-50 months |
| $T_i^{rs}$ | A variable with CDF $F_i^{rs}(t)$ denoting the time of the $i^{th}$ active MSF from L to H state. (MSF Failover time) | G | 1-3.5 seconds |
| $T_i^{rv}$ | A variable with CDF $F_i^{rv}(t)$ denoting the time of the $i^{th}$ active VM from L to H state. (VM Failover time) | G | 4-5 seconds |
| $T_i^{rm}$ | A variable with CDF $F_i^{rm}(t)$ denoting the time of the $i^{th}$ active VMM from M to H state. (VM Migration time) | G | 8-10 seconds |
| $T_i^{rbs}$ | A variable with CDF $F_i^{rbs}(t)$ denoting the time of the $i^{th}$ backup MSF from D to BR state. (Backup MSF restart time) | G | 1-3.5 seconds |
| $T_i^{rbv}$ | A variable with CDF $F_i^{rbv}(t)$ denoting the time of the $i^{th}$ backup VM from D to BR state. (Backup VM restart time) | G | 3-6 seconds |
| $T_i^{rbm}$ | A variable with CDF $F_i^{rbm}(t)$ denoting the time of the $i^{th}$ backup VMM from D to BR state. (Backup VMM reboot time) | G | 7-11 seconds |
| $T_i^{frbs}$ | A variable with CDF $F_i^{frbs}(t)$ denoting the time of the $i^{th}$ backup MSF from F to BC state. (Backup MSF fix and restart time) | G | 4-8.5 seconds |
| $T_i^{frbv}$ | A variable with CDF $F_i^{frbv}(t)$ denoting the time of the $i^{th}$ backup VM from F to BC state. (Backup VM fix and restart time) | G | 6.5-11 seconds |
| $T_i^{frbm}$ | A variable with CDF $F_i^{frbm}(t)$ denoting the time of the $i^{th}$ backup VMM from F to BC state. (Backup VMM fix and reboot time) | G | 9-13.5 seconds |
| $T_i^{rtsh}, T_i^{rtsr}, T_i^{rtsc}$ | The variables with CDFs $F_i^{rtsh}(t)=u(t-a_i^{s})$, $F_i^{rtsr}(t)=u(t-a_i^{s})$ and $F_i^{rtsc}(t)=u(t-a_i^{s})$ denoting the time of the $i^{th}$ active MSF from D to L state under the corresponding backup MSF at H, BR and BC state, respectively. (RTI) | U | 0-1800 minutes |
| $T_i^{rtvh}, T_i^{rtvr}, T_i^{rtvc}$ | The variables with CDFs $F_i^{rtvh}(t)=u(t-a_i^{v})$, $F_i^{rtvr}(t)=u(t-a_i^{v})$ and $F_i^{rtvc}(t)=u(t-a_i^{v})$ denoting the time of the $i^{th}$ active VM from D to L state under the corresponding backup VM at H, BR and BC state, respectively. (RTI) | U | 0-3600 minutes |
| $T_i^{rtmh}, T_i^{rtmr}, T_i^{rtmc}$ | The variables with CDFs $F_i^{rtmh}(t)=u(t-a_i^{m})$, $F_i^{rtmr}(t)=u(t-a_i^{m})$ and $F_i^{rtmc}(t)=u(t-a_i^{m})$ denoting the time of the $i^{th}$ active VMM from D to M state under the corresponding backup VMM at H, BR and BC state, respectively. (RTI) | U | 0-7200 minutes |
| $T_i^{RV}$ | A variable with CDF $F_i^{RV}(t)$ denoting the time of all active MSFs, backup MSFs, active VMs and backup VMs in the $i^{th}$ Primary Metaverse host and its Backup Metaverse host from R to H state. (Time of restarting all SFs and VMs) | G | 0.1-0.95 minutes |
| $T_i^{RM}$ | A variable with CDF $F_i^{RM}(t)$ denoting the time of all active MSFs, backup MSFs, active VMs, backup VMs, active VMMs and backup VMMs in the $i^{th}$ Primary Metaverse host and its Backup Metaverse host from R to H state. (Time of restarting/rebooting all MSFs, VMs and VMMs) | G | 0.5-1.05 minutes |
| $T_i^{R}$ | A variable with CDF $F_i^{R}(t)$ denoting the time of the $i^{th}$ Primary Metaverse host and its corresponding Backup Metaverse host from failure to robustness. (Host fix time) | G | 0.1-0.35 hours |
| $T_i^{asvh}$ | A variable with CDF $F_i^{asvh}(t)$ denoting the minimum time of active MSF and VM in the $i^{th}$ Primary Metaverse host and its Backup Metaverse host from H to D state after the active VMM suffering from software aging. (Aging time) | EXP | -- |

'--' represents that the default value depends on the other variables. 'Dis.' represents distribution. 'EXP' represents exponential distribution. 'G' represents general distribution. 'U' represents unit step function.

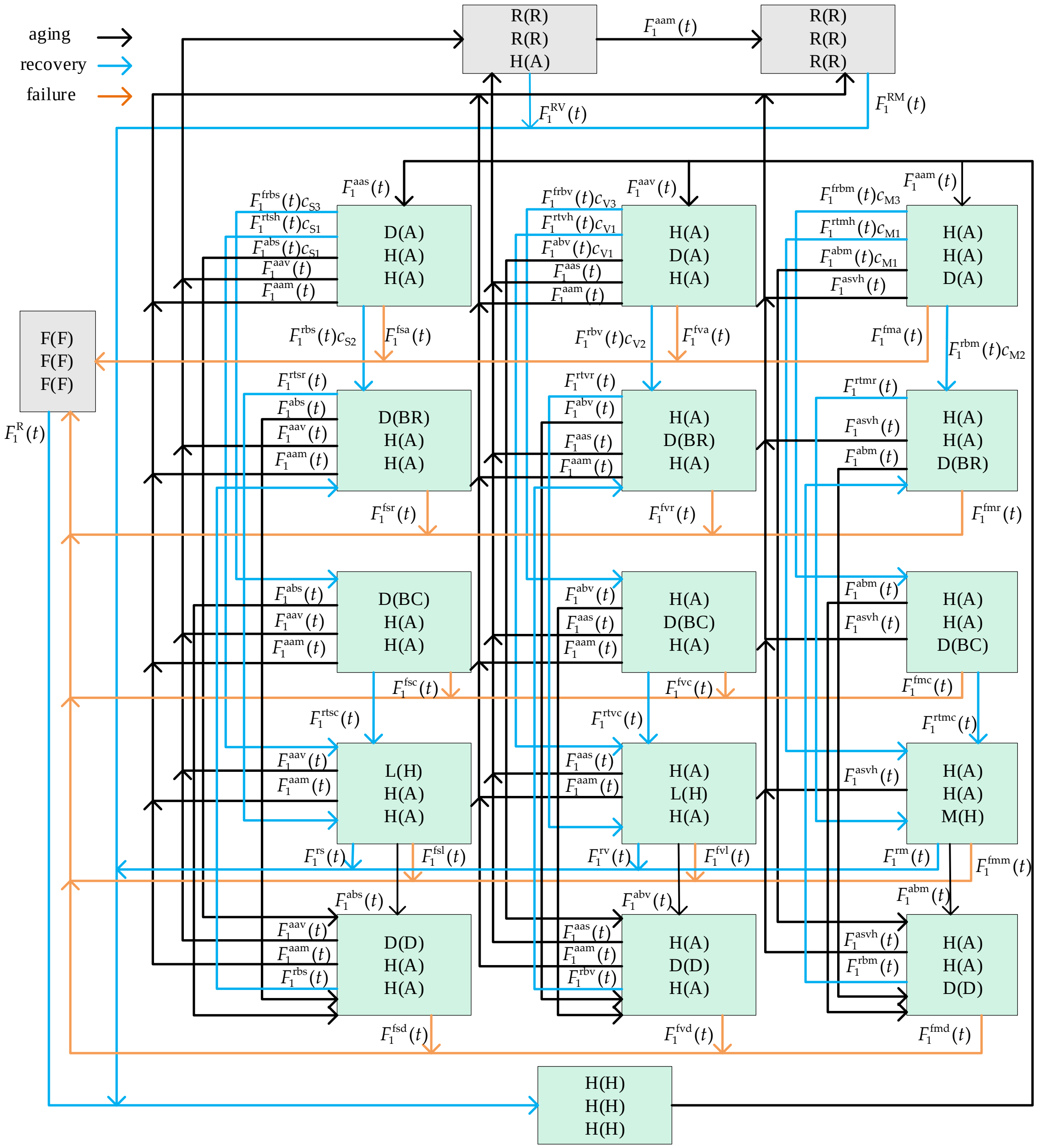

Fig. 3 SMP sub-model for capturing the behaviors of a Primary Metaverse host and its Backup Metaverse host

- Backup component suffering from failure. This backup component will be fixed and then restarted to support rejuvenation technique. After restarting the backup component and waiting for a while, rejuvenation technique will be triggered.

Backup component can still be affected by software aging during the rejuvenation technique-triggered intervals or during the rejuvenation technique execution. Once this situation happens, the backup component will be restarted. In addition, there are four cases that can happen before the backup component takes over the work of the active component:

- Before the backup MSF completes the failover, the active VM responsible for running its corresponding active MSF can experience software aging. Before the backup VM completes the failover, the active MSF running on its corresponding active VM can experience software aging. In both cases, the active MSF, backup MSF, this active VM and its backup VM will be restarted.
- Before the backup MSF/VM completes the failover, the active VMM can experience software aging. Before the VM migration technique is completed, the active MSF or active VM in a Primary Metaverse host can experience software aging. In both cases, all components in this Primary Metaverse host and its backup Metaverse host will be restarted/rebooted.

Then, based on the above analysis, we can study the behaviors of serial and parallel MSFC systems:

- In a serial MSFC system, if any component crashes, the request processing stops. The serial MSFC system becomes unavailable.
- If one of the parallel components crashes, the parallel MSFC system is still available because requests can be processed as long as there are still functional components in the parallel part of the parallel MSFC system. However, if any serial component in the parallel MSFC system crashes, service also will become unavailable.

We define that if an MSFC system crashes, all MSFs, VMs, and VMMs in this system will be restarted/rebooted in sequence after the failed active component finishes its fixing.

## 3.2 System Model

This section introduces the hierarchical model for evaluating the dependability of MSFC system. The proposed hierarchical model includes two levels: SMP sub-model and RBD sub-model.

### 3.2.1 SMP Sub-model

Fig. 3 gives the SMP sub-model. In the boxes of Fig. 3, from top to bottom, the first row represents the active MSF and its backup, the second row represents the active VM and its backup, and the third row represents the active VMM and its backup. Furthermore, the letters outside the parentheses indicate the states of active components, while the letters inside the parentheses denote the states of their corresponding backup components. There are nine component states: Healthy, Degradation, Restart/Reboot, Failover, Migration, Failed, Arbitrary, Backup Component Restart/Reboot and Backup Component Recovery, denoted by H, D, R, L, M, F, A, BR and BC, respectively. The detailed meanings are given as follows:

- State H (Healthy): In this state, the component is robust and the requests can be processed normally. The component suffering from software aging or failure can return to this state through recovery operations.
- State D (Degradation): In this state, the component is still working, but at a low-efficient execution phase. The ability of this component to provide services in this state is lower than that of components in Healthy state.
- State R (Restart/Reboot): In this state, the component will be restarted/rebooted.
- State L (Failover): In this state, failover technique is triggered and the backup component is ready to take over the work.
- State M (Migration): In this state, VM migration technique is triggered and backup VMM will take over the work.
- State F (Failed): In this state, the component fails due to software aging. All components must be restarted/rebooted after fixing the failed component to back to the Healthy state.
- State A (Arbitrary): In this state, it is unknown whether the backup component is healthy. The probabilities of these three cases occurring are $C_{i1}, C_{i2}, C_{i3}$, respectively. Here, ($C_{i1}+C_{i2}+C_{i3}=1$)
- State BR (Backup Component Restart/Reboot): In this state, the restarting/rebooting of backup component is complete.
- State BC (Backup Component Recovery): In this state, the fixing and restarting/rebooting of backup component is complete.

Active component has five states, namely H, D, R, L/M and F. Backup component has seven states, namely H, D, R, F, A, BR and BC. Based on the description above, we describe the state of a Primary Metaverse host and its Backup Metaverse host by defining a 6-tuple index $(i_{a_i}^{sf}, i_{b_i}^{sf}, j_{a_i}^{vm}, j_{b_i}^{vm}, k_{a_i}^{vmm}, k_{b_i}^{vmm})$. $i_{a_i}^{sf}$ and $i_{b_i}^{sf}$ are the states of the $i^{th}$ active MSF and backup MSF, respectively. $j_{a_i}^{vm}$ and $j_{b_i}^{vm}$ are the states of the $i^{th}$ active VM and backup VM, respectively. $k_{a_i}^{vmm}$ and $k_{b_i}^{vmm}$ are the states of the $i^{th}$ active VMM and backup VMM, respectively. Therefore, there are total $5^3 * 7^3$ states, where $5^3 * 7^3 - 19$ states are meaningless and can be ignored.

We assume that the aging event-occurring times are exponentially distributed. Other event-occurring times follow non-exponential distribution.

### 3.2.2 RBD Sub-model

We model the serial and parallel MSFC as RBD (Fig. 4). By using RBD, we can better analyze the behaviors of an MSFC consisting of multiple Primary Metaverse hosts and their Backup Metaverse hosts, and better understand how the behaviors of different components in the serial and parallel parts affects the MSFC dependability.

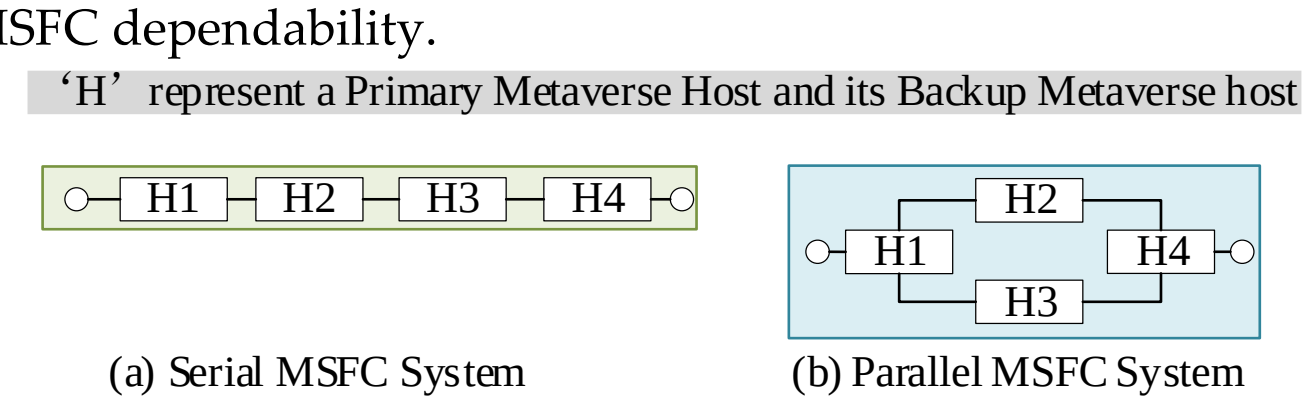

Fig. 4 RBD sub-model for MSFC systems

## 3.3 System Availability Analysis

This section presents formulas for calculating the steady-state availability of an MSFC consisting of arbitrary number of MSFs. The details are as follows.

At first, we give the process to calculate the MSF steady-state availability.

Fig. 5 shows the kernel matrix $\mathbf{K}(t)$. In this kernel matrix, the non-zero element $k_{s_i s_j}(t)$ represents the conditional probability that the MSF has in state $s_i$ and will enter state $s_j$ at time *t*. The formulas for calculating the non-zero elements in this kernel matrix are given in Section A of the supplementary file.

Then, we construct the one-step transition probability matrix (TPM) $\mathbf{P}=[p_{s_i s_j}]$ to describe the embedded discrete-time Markov chain (EDTMC) of the SMP sub-model. $\mathbf{P}=\lim_{t\to\infty}\mathbf{K}(t)$ are given in Section A of the supplementary file.

Next, the steady-state probability vector $\mathbf{V}=[\mathbf{V}_{s_i}]$ of the EDTMC can be obtained by Equation (1). Here, $e$ denotes the column vector where all entries are equal to 1.

$$\mathbf{V}=\mathbf{VP} \text{ subject to } \mathbf{V}e^T=1 \tag{1}$$

We give the calculation formula of $\mathbf{V}_{s_i}$ in Section A of the supplementary file.

Then, the mean sojourn time $h_{s_i}$ in state $s_i$ can be calculated according to Equation (2).

$$h_{s_i}=\int_0^{\infty}(1-G_i(t))dt \tag{2}$$

$$
\mathbf{K}(\mathrm{t})=\begin{pmatrix}
0 & 0 & 0 & 0 & k_{s_0s_4}(t) & 0 & 0 & 0 & 0 & k_{s_0s_9}(t) & 0 & 0 & 0 & 0 & k_{s_0s_{14}}(t) & 0 & 0 & 0 & 0 \\
k_{s_1s_0}(t) & 0 & 0 & 0 & 0 & 0 & 0 & 0 & 0 & 0 & 0 & 0 & 0 & 0 & 0 & 0 & 0 & 0 & 0 \\
k_{s_2s_0}(t) & 0 & 0 & k_{s_2s_3}(t) & 0 & 0 & 0 & 0 & 0 & 0 & 0 & 0 & 0 & 0 & 0 & 0 & 0 & 0 & 0 \\
k_{s_3s_0}(t) & 0 & 0 & 0 & 0 & 0 & 0 & 0 & 0 & 0 & 0 & 0 & 0 & 0 & 0 & 0 & 0 & 0 & 0 \\
0 & k_{s_4s_1}(t) & k_{s_4s_2}(t) & k_{s_4s_3}(t) & 0 & k_{s_4s_5}(t) & k_{s_4s_6}(t) & k_{s_4s_7}(t) & k_{s_4s_8}(t) & 0 & 0 & 0 & 0 & 0 & 0 & 0 & 0 & 0 & 0 \\
0 & k_{s_5s_1}(t) & k_{s_5s_2}(t) & k_{s_5s_3}(t) & 0 & 0 & 0 & k_{s_5s_7}(t) & k_{s_5s_8}(t) & 0 & 0 & 0 & 0 & 0 & 0 & 0 & 0 & 0 & 0 \\
0 & k_{s_6s_1}(t) & k_{s_6s_2}(t) & k_{s_6s_3}(t) & 0 & 0 & 0 & k_{s_6s_7}(t) & k_{s_6s_8}(t) & 0 & 0 & 0 & 0 & 0 & 0 & 0 & 0 & 0 & 0 \\
k_{s_7s_0}(t) & k_{s_7s_1}(t) & k_{s_7s_2}(t) & k_{s_7s_3}(t) & 0 & 0 & 0 & 0 & k_{s_7s_8}(t) & 0 & 0 & 0 & 0 & 0 & 0 & 0 & 0 & 0 & 0 \\
0 & k_{s_8s_1}(t) & k_{s_8s_2}(t) & k_{s_8s_3}(t) & 0 & k_{s_8s_5}(t) & 0 & 0 & 0 & 0 & 0 & 0 & 0 & 0 & 0 & 0 & 0 & 0 & 0 \\
0 & k_{s_9s_1}(t) & k_{s_9s_2}(t) & k_{s_9s_3}(t) & 0 & 0 & 0 & 0 & 0 & 0 & k_{s_9s_{10}}(t) & k_{s_9s_{11}}(t) & k_{s_9s_{12}}(t) & k_{s_9s_{13}}(t) & 0 & 0 & 0 & 0 & 0 \\
0 & k_{s_{10}s_1}(t) & k_{s_{10}s_2}(t) & k_{s_{10}s_3}(t) & 0 & 0 & 0 & 0 & 0 & 0 & 0 & 0 & k_{s_{10}s_{12}}(t) & k_{s_{10}s_{13}}(t) & 0 & 0 & 0 & 0 & 0 \\
0 & k_{s_{11}s_1}(t) & k_{s_{11}s_2}(t) & k_{s_{11}s_3}(t) & 0 & 0 & 0 & 0 & 0 & 0 & 0 & 0 & k_{s_{11}s_{12}}(t) & k_{s_{11}s_{13}}(t) & 0 & 0 & 0 & 0 & 0 \\
k_{s_{12}s_0}(t) & k_{s_{12}s_1}(t) & k_{s_{12}s_2}(t) & k_{s_{12}s_3}(t) & 0 & 0 & 0 & 0 & 0 & 0 & 0 & 0 & 0 & k_{s_{12}s_{13}}(t) & 0 & 0 & 0 & 0 & 0 \\
0 & k_{s_{13}s_1}(t) & k_{s_{13}s_2}(t) & k_{s_{13}s_3}(t) & 0 & 0 & 0 & 0 & 0 & 0 & k_{s_{13}s_{10}}(t) & 0 & 0 & 0 & 0 & 0 & 0 & 0 & 0 \\
0 & k_{s_{14}s_1}(t) & 0 & k_{s_{14}s_3}(t) & 0 & 0 & 0 & 0 & 0 & 0 & 0 & 0 & 0 & 0 & 0 & k_{s_{14}s_{15}}(t) & k_{s_{14}s_{16}}(t) & k_{s_{14}s_{17}}(t) & k_{s_{14}s_{18}}(t) \\
0 & k_{s_{15}s_1}(t) & 0 & k_{s_{15}s_3}(t) & 0 & 0 & 0 & 0 & 0 & 0 & 0 & 0 & 0 & 0 & 0 & 0 & 0 & k_{s_{15}s_{17}}(t) & k_{s_{15}s_{18}}(t) \\
0 & k_{s_{16}s_1}(t) & 0 & k_{s_{16}s_3}(t) & 0 & 0 & 0 & 0 & 0 & 0 & 0 & 0 & 0 & 0 & 0 & 0 & 0 & k_{s_{16}s_{17}}(t) & k_{s_{16}s_{18}}(t) \\
k_{s_{17}s_0}(t) & k_{s_{17}s_1}(t) & 0 & k_{s_{17}s_3}(t) & 0 & 0 & 0 & 0 & 0 & 0 & 0 & 0 & 0 & 0 & 0 & 0 & 0 & 0 & k_{s_{17}s_{18}}(t) \\
0 & k_{s_{18}s_1}(t) & 0 & k_{s_{18}s_3}(t) & 0 & 0 & 0 & 0 & 0 & 0 & 0 & 0 & 0 & 0 & 0 & k_{s_{18}s_{15}}(t) & 0 & 0 & 0
\end{pmatrix}
$$

Fig. 5 Kernel matrix $\mathbf{K}(\mathrm{t})$

where the distribution function $G_i(t)$ is used to represent the sojourn time distribution in state $s_i$. We give the calculation formula of the mean sojourn time $h_{s_i}$ in Section A of the supplementary file.

We then can obtain MSF steady-state availability $\pi_{s_i}$ at state $s_i$ according to Equation (3).

$$\pi_{s_i}=\frac{V_{s_i}h_{s_i}}{\sum_{s_j}V_{s_j}h_{s_j}} \tag{3}$$

where $V_{s_i}$ and $h_{s_i}$ can be gained from Equations (1) and (2).

Finally, the steady-state availability of MSF $\mathrm{A}_w$ is computed by Equation (4).

$$\mathrm{A}_w=1-(\pi_{s_1}+\pi_{s_2}+\pi_{s_3}) \tag{4}$$

where $\pi_{s_1}, \pi_{s_2}$, and $\pi_{s_3}$ are MSF steady-state availability in states $s_1$, $s_2$ and $s_3$, respectively.

Then, we give the formulas of calculating the steady-state availability of MSFC.

For a serial MSFC with $n$ MSFs, the calculation formula of steady-state availability $\mathrm{A}_s$ is shown in Equation (5), where $\mathrm{A}_w$ is the steady-state availability of the $w^{th}$ MSF. And for a parallel MSFC with $m$ serial MSFs and $n$-$m$ parallel MSFs, the calculation formula of the steady-state availability $\mathrm{A}_p$ is shown in Equation (6).

$$\mathrm{A}_s=\prod_{w=1}^{n}\mathrm{A}_w \tag{5}$$

$$\mathrm{A}_p=(1-\prod_{w=m+1}^{n}(1-\mathrm{A}_w))\prod_{w=1}^{m}\mathrm{A}_w \tag{6}$$

### 3.4 System Reliability Analysis

This section presents the formulas for calculating the reliability of MSFC consisting of $n$ MSFs. The classical reliability assessment metric, MTTF, is calculated in a scenario where recovery operations are not performed after MSFC system failure.

At first, we give the calculation process of MSF MTTF.

We construct a kernel matrix $\mathbf{K}'(\mathrm{t})$, which is deformed from $\mathbf{K}(\mathrm{t})$ in Section 3.3. The formulas for calculating the non-zero elements in this kernel matrix are given in Section B of the supplementary file.

Then we can obtain the one-step TPM $\mathbf{P}'$ describing the EDTMC in the SMP sub-model with absorbing states by the method described in the previous Section 3.3. $\mathbf{P}'$ and the formulas for calculating the non-zero elements in it are given in Section B of the supplementary file.

The expected number of visits $V_{s_{i'}}^{*}$ to state $s_{i'}$ until absorption is calculated by applying Equation (7).

$$V_{s_{i'}}^{*}=\alpha_{s_{i'}}+\sum_{j'=0}^{15}V_{s_{j'}}^{*}p_{s_{j'}s_{i'}} \tag{7}$$

where $\alpha_{s_{i'}}$ is the initial probability in state $s_{i'}$. $V_{s_0}^{'}$ and $V_{s_{i'}}^{'}$ ($1\le i'\le 15$) are given by Equation (8) and (9), respectively.

$$V_{s_0}^{'}=(-1)\Big/(\sum_{i'=1}^{15}p_{s_0s_{i'}}^{'}p_{s_{i'}s_0}^{'}-1) \tag{8}$$

$$V_{s_{i'}}^{'}=(-p_{s_0s_{i'}}^{'})\Big/(\sum_{i'=1}^{15}p_{s_0s_{i'}}^{'}p_{s_{i'}s_0}^{'}-1) \tag{9}$$

The mean sojourn time $h_{s_{i'}}^{'}$, in state $s_{i'}$ is given in Section B of the supplementary file.

Finally, MTTF of the $w^{th}$ MSF, $\mathrm{MTTF}_w$, is computed by Equation (10).

$$\mathrm{MTTF}_w=\sum_{i'=0}^{15}V_{s_{i'}}^{'}h_{s_{i'}}^{'}\,(w=1,2,3,\ldots) \tag{10}$$

where $V_{s_{i'}}^{'}$ and $h_{s_{i'}}^{'}$ can be obtained from Equation (8), (9)and Section B of the supplementary file.

Then, we give the formulas of calculating the MTTF of MSFC.

For a serial MSFC with $n$ MSFs, the calculation formulas of MTTF ($\mathrm{MTTF}_s$) is shown in Equation (11). And for a parallel MSFC with $m$ serial MSFs and $n$-$m$ parallel MSFs, the formulas of calculating MTTF ($\mathrm{MTTF}_p$) is shown in Equation (12).

$$\mathrm{MTTF}_s=\min(\mathrm{MTTF}_1,\ldots,\mathrm{MTTF}_w,\ldots,\mathrm{MTTF}_n) \tag{11}$$

$$\mathrm{MTTF}_p=\min(\mathrm{MTTF}_1,\ldots,\mathrm{MTTF}_w,\ldots,\mathrm{MTTF}_m,\max(\mathrm{MTTF}_{m+1},\mathrm{MTTF}_{m+2},\ldots,\mathrm{MTTF}_n)) \tag{12}$$

## 4 EXPERIMENTAL EVALUATION

This section first conducts simulation experiments to verify our proposed model and formulas. Then, we conduct sensitivity analysis experiments and further conduct numerical analysis experiments for key parameters. Finally, we analyze the effects of MSF number, RTIs, backup components' behaviors and CDF types on the MSFC steady-state availability and MTTF.

### 4.1 Experiment Configuration

TABLE 2, TABLE 3, and TABLE 4 provide the default parameter settings and the CDF types for event-occurring times in the experiments. These settings and CDF types are utilized to showcase the effectiveness of our model and formulas. The default values of parameters are obtained from prior literature [20]. Additionally, our model also applies other parameter settings and CDF types. The simulation and numerical experiments are performed in MAPLE [30].

TABLE 4
CDF TYPE OF EVENT-OCCURRING TIMES IN EXPERIMENTS

| CDF | Type | Para. | CDF | Type | Para. |
|---|---|---|---|---|---|
| $F_i^{\text{aas}}(t)$ | EXP | $\theta_i^{\text{aas}}$ | $F_i^{\text{rm}}(t)$ | EXP | $\gamma_i^{\text{rm}}$ |
| $F_i^{\text{aav}}(t)$ | EXP | $\theta_i^{\text{aav}}$ | $F_i^{\text{rbs}}(t)$ | EXP | $\eta_i^{\text{rbs}}$ |
| $F_i^{\text{aam}}(t)$ | EXP | $\theta_i^{\text{aam}}$ | $F_i^{\text{rbv}}(t)$ | EXP | $\eta_i^{\text{rbv}}$ |
| $F_i^{\text{abs}}(t)$ | EXP | $\theta_i^{\text{abs}}$ | $F_i^{\text{rbm}}(t)$ | EXP | $\eta_i^{\text{rbm}}$ |
| $F_i^{\text{abv}}(t)$ | EXP | $\theta_i^{\text{abv}}$ | $F_i^{\text{frbs}}(t)$ | EXP | $\lambda_i^{\text{frbs}}$ |
| $F_i^{\text{abm}}(t)$ | EXP | $\theta_i^{\text{abm}}$ | $F_i^{\text{frbv}}(t)$ | EXP | $\lambda_i^{\text{frbv}}$ |
| $F_i^{\text{fsa}}(t)$, $F_i^{\text{fsr}}(t)$, $F_i^{\text{fsc}}(t)$, $F_i^{\text{fsd}}(t)$ | HYPO | $\alpha_{i1}^{\text{fsa}}, \alpha_{i2}^{\text{fsa}}$ $\alpha_{i1}^{\text{fsr}}, \alpha_{i2}^{\text{fsr}}$ $\alpha_{i1}^{\text{fsc}}, \alpha_{i2}^{\text{fsc}}$ $\alpha_{i1}^{\text{fsd}}, \alpha_{i2}^{\text{fsd}}$ | $F_i^{\text{rtsh}}(t)$, $F_i^{\text{rtsr}}(t)$, $F_i^{\text{rtsc}}(t)$ | U | $\omega_i^{\text{s}}$ |
| $F_i^{\text{fsl}}(t)$ | HYPO | $\alpha_{i1}^{\text{fsl}}, \alpha_{i2}^{\text{fsl}}$ | $F_i^{\text{frbm}}(t)$ | EXP | $\lambda_i^{\text{frbm}}$ |
| $F_i^{\text{fva}}(t)$, $F_i^{\text{fvr}}(t)$, $F_i^{\text{fvc}}(t)$, $F_i^{\text{fvd}}(t)$ | HYPO | $\beta_{i1}^{\text{fva}}, \beta_{i2}^{\text{fva}}$ $\beta_{i1}^{\text{fvr}}, \beta_{i2}^{\text{fvr}}$ $\beta_{i1}^{\text{fvc}}, \beta_{i2}^{\text{fvc}}$ $\beta_{i1}^{\text{fvd}}, \beta_{i2}^{\text{fvd}}$ | $F_i^{\text{rtvh}}(t)$, $F_i^{\text{rtvr}}(t)$, $F_i^{\text{rtvc}}(t)$ | U | $\omega_i^{\text{v}}$ |
| $F_i^{\text{fvl}}(t)$ | HYPO | $\beta_{i1}^{\text{fvl}}, \beta_{i2}^{\text{fvl}}$ | $F_i^{\text{RV}}(t)$ | EXP | $\mu_i^{\text{RV}}$ |
| $F_i^{\text{fma}}(t)$, $F_i^{\text{fmr}}(t)$, $F_i^{\text{fmc}}(t)$, $F_i^{\text{fmd}}(t)$ | HYPO | $\delta_{i1}^{\text{fma}}, \delta_{i2}^{\text{fma}}$ $\delta_{i1}^{\text{fmr}}, \delta_{i2}^{\text{fmr}}$ $\delta_{i1}^{\text{fmc}}, \delta_{i2}^{\text{fmc}}$ $\delta_{i1}^{\text{fmd}}, \delta_{i2}^{\text{fmd}}$ | $F_i^{\text{rtmh}}(t)$, $F_i^{\text{rtmr}}(t)$, $F_i^{\text{rtmc}}(t)$ | U | $\omega_i^{\text{m}}$ |
| $F_i^{\text{fmm}}(t)$ | HYPO | $\delta_{i1}^{\text{fmm}}, \delta_{i2}^{\text{fmm}}$ | $F_i^{\text{RM}}(t)$ | EXP | $\mu_i^{\text{RM}}$ |
| $F_i^{\text{rs}}(t)$ | EXP | $\gamma_i^{\text{rs}}$ | $F_i^{\text{R}}(t)$ | EXP | $\mu_i^{\text{R}}$ |
| $F_i^{\text{rv}}(t)$ | EXP | $\gamma_i^{\text{rv}}$ | $F_i^{\text{asvh}}(t)$ | EXP | $\kappa_i^{\text{asvh}}$ |

'Para.' represents Parameter. 'EXP', 'HYPO' and 'U' represent exponential distribution, hypoexponential distribution and unit step function, respectively.

### 4.2 Verification of Our Model and Formulas

In this section, we will verify the proposed model and formula in terms of steady-state availability and MTTF by comparing simulation and numerical results for MSF, serial MSFC and parallel MSFC. We choose the parameter $\mu_1^{\text{R}}$ for the verification experiment of steady-state availability and the parameter $\alpha_1^{\text{fsa}}$ for the verification experiment of MTTF. The reasons for choosing these two parameters are given in TABLE 5 in Section 4.3, which have the greatest impact on steady-state availability and MTTF, respectively. Fig. 6 and Fig. 7 show the comparison between the simulation and numerical results for MSF steady-state availability and MTTF, respectively. Fig. 8 and Fig. 9 show the comparison between the simulation and numerical results for serial MSFC steady-state availability and MTTF, respectively. In these figures, 'Sim' and 'Num' denote the simulation and numerical results, respectively. The simulation and numerical results of parallel MSFC steady-state availability and MTTF are given in Section C of the supplementary file. From the experimental results, it can be seen that the simulation results of MSF, serial MSFC or parallel MSFC dependability are very close to the numerical results. These results show that our models and formulas are approximately accurate.

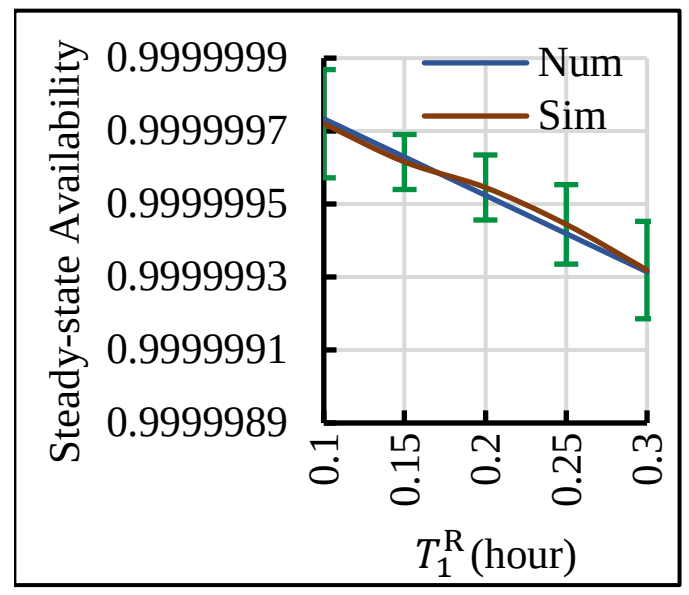

Fig. 6 Comparison of simulation and numerical results for MSF steady-state availability

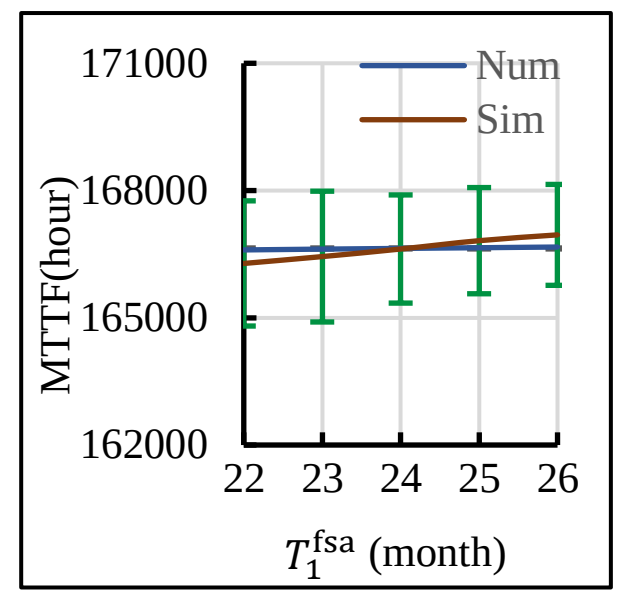

Fig. 7 Comparison of simulation and numerical results for MSF MTTF

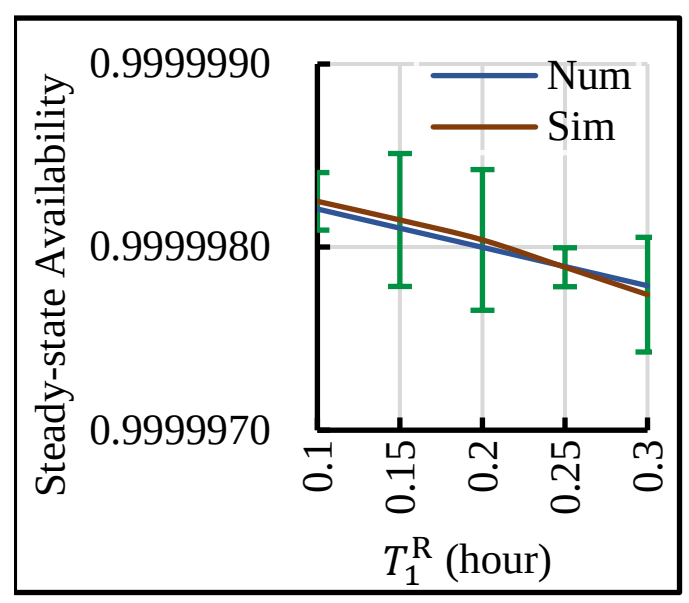

Fig. 8 Comparison of simulation and numerical results for steady-state availability of serial MSFC system

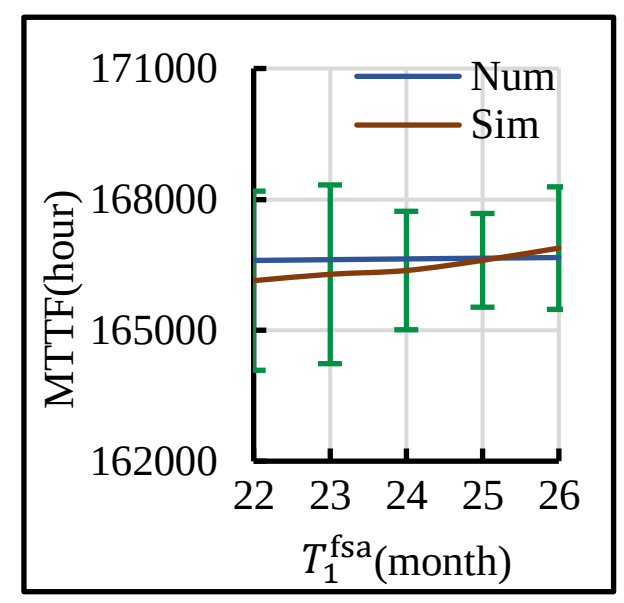

Fig. 9 Comparison of simulation and numerical results for MTTF of serial MSFC system

### 4.3 Sensitivity Analysis

In this section, we analyze the sensitivity of steady-state availability and MTTF of the serial and parallel MSFC with respect to different parameters. We calculate the scaled sensitivity by using Equation (13).

$$SS_{\rho}(Y) = \frac{\partial Y}{\partial \rho}\left(\frac{\rho}{Y}\right) \tag{13}$$

where $Y$ is the calculation formula for each dependability measure we derive in Section 3. TABLE 5 shows the sensitivities of serial and parallel MSFC dependability measures with respect to the system parameters. We can observe that:

- A positive sensitivity indicates that an increase in parameter value leads to an increase in metrics of interest, and a negative sensitivity indicates that an increase in parameter value leads to a decrease in metrics of interest. This is because an increase in the aging time and failure time leads to an increase in the available time of the MSFC system, thereby enhancing the MSFC dependability. An increase in the recovery time results in an increasing probability of component failure due to software aging during the recovery process, or an increased probability of the aging of other components, leading to a decrease in the MSFC dependability.

- Among the parameters calculated, $\mu_1^{\mathrm{R}}$ (related to the time of the 1st Primary Metaverse host and its Backup Metaverse host from failure to robustness) and $\mu_1^{\mathrm{RM}}$ (related to the time of restarting/rebooting all MSFs, VMs and VMMs in the 1st Primary Metaverse host and its Backup Metaverse host) are the first and the second most important parameters influencing the steady-state availability, respectively. On the other hand, $\alpha_1^{\mathrm{fsa}}$ (related to the 1st active MSF failure time when backup MSF at arbitrary state) is the most important parameter affecting the MTTF. The MSFC dependability optimization focuses on identifying these crucial parameters that have the significant influence on dependability.

TABLE 5
STEADY-STATE AVAILABILITY AND MTTF SENSITIVITY

| $\rho$ | Serial MSFC system | | Parallel MSFC system | |
|---|---|---|---|---|
| | $SS_\rho(\pi_{\mathrm{A}})$ | $SS_\rho(\mathrm{MTTF})$ | $SS_\rho(\pi_{\mathrm{A}})$ | $SS_\rho(\mathrm{MTTF})$ |
| $\boldsymbol{\alpha_1^{\mathrm{fsa}}}$ | -1.46E-09 | -3.08E-03 | -7.40E-16 | -2.98E-03 |
| $\alpha_1^{\mathrm{fsr}}$ | -9.78E-20 | -2.03E-13 | -4.97E-26 | -1.94E-13 |
| $\alpha_1^{\mathrm{fsc}}$ | -2.25E-21 | -4.62E-15 | -1.15E-27 | -4.44E-15 |
| $\alpha_1^{\mathrm{fsd}}$ | -2.35E-29 | -5.09E-23 | -1.20E-35 | -4.80E-23 |
| $\alpha_1^{\mathrm{fsl}}$ | -3.71E-27 | -2.03E-12 | -1.89E-33 | -2.01E-12 |
| $\beta_1^{\mathrm{fva}}$ | -6.77E-11 | -1.11E-04 | -3.44E-17 | -9.60E-05 |
| $\beta_1^{\mathrm{fvr}}$ | -7.74E-19 | -1.21E-12 | -3.93E-25 | -1.18E-12 |
| $\beta_1^{\mathrm{fvc}}$ | -7.16E-21 | -1.17E-14 | -3.64E-27 | -1.15E-14 |
| $\beta_1^{\mathrm{fvd}}$ | -2.15E-28 | -2.74E-22 | -1.09E-34 | -2.66E-22 |
| $\beta_1^{\mathrm{fvl}}$ | -1.11E-26 | -2.49E-12 | -5.67E-33 | -2.40E-12 |
| $\delta_1^{\mathrm{fma}}$ | -1.24E-09 | -2.56E-03 | -6.31E-16 | -2.54E-03 |
| $\delta_1^{\mathrm{fmr}}$ | -6.29E-19 | -1.30E-12 | -3.20E-25 | -1.27E-12 |
| $\delta_1^{\mathrm{fmc}}$ | -5.62E-21 | -1.15E-14 | -2.86E-27 | -1.12E-14 |
| $\delta_1^{\mathrm{fmd}}$ | -2.95E-28 | -6.44E-22 | -1.50E-34 | -6.28E-22 |
| $\delta_1^{\mathrm{fmm}}$ | -2.04E-26 | -2.80E-12 | -1.04E-32 | -2.60E-12 |
| $\gamma_1^{\mathrm{rs}}$ | -1.63E-14 | 7.63E-07 | -8.30E-21 | 7.29E-07 |
| $\gamma_1^{\mathrm{rv}}$ | -3.11E-14 | 1.18E-06 | -1.58E-20 | 1.14E-06 |
| $\gamma_1^{\mathrm{rm}}$ | -4.18E-14 | 1.95E-06 | -2.13E-20 | 1.89E-06 |
| $\eta_1^{\mathrm{rbs}}$ | -6.32E-16 | -1.31E-09 | -3.21E-22 | -1.25E-09 |
| $\eta_1^{\mathrm{rbv}}$ | -3.18E-15 | -4.96E-09 | -1.62E-21 | -4.79E-09 |
| $\lambda_1^{\mathrm{frbs}}$ | -4.52E-17 | -9.27E-11 | -2.30E-23 | -8.91E-11 |
| $\lambda_1^{\mathrm{frbv}}$ | -6.09E-17 | -9.98E-11 | -3.09E-23 | -9.64E-11 |
| $\lambda_1^{\mathrm{frbm}}$ | -5.20E-17 | -1.07E-10 | -2.64E-23 | -8.68E-11 |
| $\mu_1^{\mathrm{RV}}$ | 1.66E-08 | -- | 8.42E-15 | -- |
| $\boldsymbol{\mu_1^{\mathrm{RM}}}$ | 4.01E-08 | -- | 2.04E-14 | -- |
| $\boldsymbol{\mu_1^{\mathrm{R}}}$ | 4.72E-07 | -- | 2.40E-13 | -- |

'-' indicates that this parameter does not affect MTTF of the MSFC system

Subsequently, we conduct numerical experiments based on sensitivity analysis to establish the necessary range of critical parameters for ensuring availability. Fig. 10 shows the steady-state availability of serial MSFC over $T_1^{\mathrm{aas}}$ (the 1st active MSF aging time) and $T_1^{\mathrm{R}}$ (the 1st Primary Metaverse host and its Backup Metaverse host fix time). We observe that when the $T_1^{\mathrm{aas}}$ is 24 months and the $T_1^{\mathrm{R}}$ increases from 0.1 hours to 0.35 hours, the steady-state availability of serial MSFC decreases from 0.999998208225559 to 0.999997684076184. The steady-state availability of parallel MSFC over $T_1^{\mathrm{aas}}$ and $T_1^{\mathrm{R}}$ are given in Section C of the supplementary file.

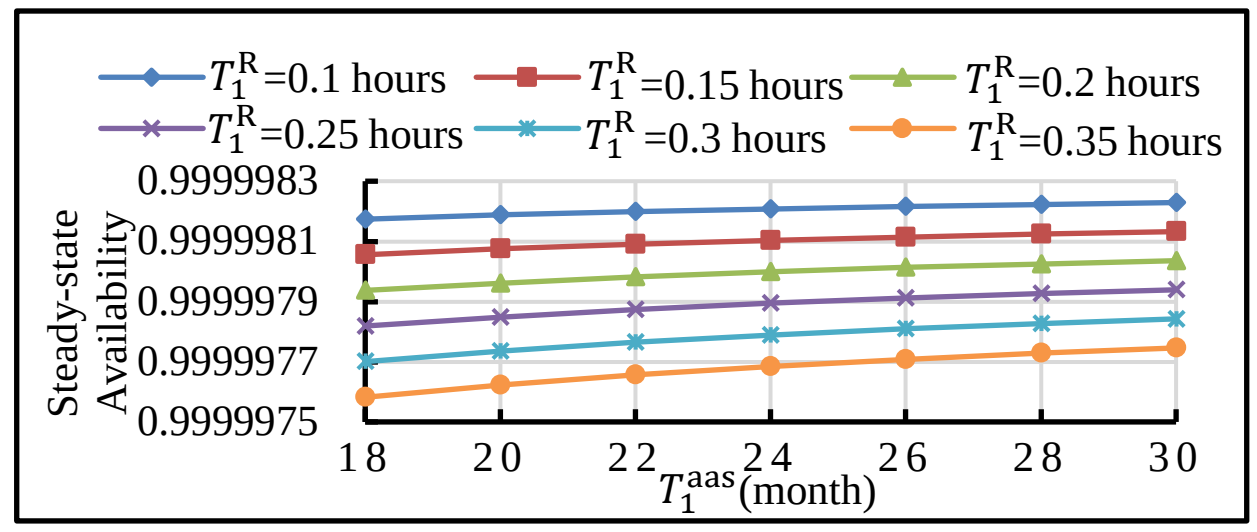

Fig. 10 Steady-state availability of serial MSFC over different $T_1^{\mathrm{aas}}$ and $T_1^{\mathrm{R}}$

## 4.4 Effect of the Number of MSFs on MSFC Dependability

In this section, we investigate the effect of the number of MSFs ($n$) on steady-state availability and MTTF of **serial and parallel MSFC**. In this experiment, we take $n$=4, $n$=5, and $n$=6 as examples. The experimental results are shown in Fig. 11 and Fig. 12. It can be observed from Fig. 11 that the steady-state availability of **serial MSFC** decreases as the number of MSFs increases and the **parallel MSFC** steady-state availability increases slightly as the number of parallel MSFs increases. We can observe from Fig. 12 that for **both serial and parallel MSFCs**, the MTTF decreases as the number of MSFs increases. It can be explained that the steady-state availability of MSFC is affected by the probability of the MSFC system being at available states and MTTF is affected by the intervals between failure event-occurring times.

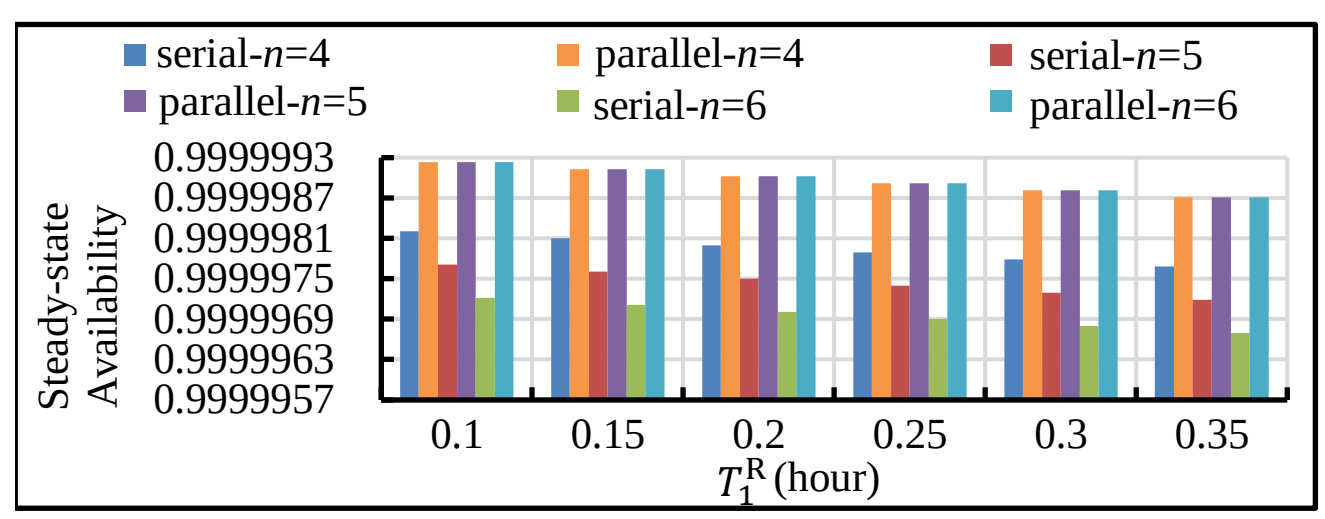

Fig. 11 The **serial and parallel MSFC steady-state availability** over the number of MSFs

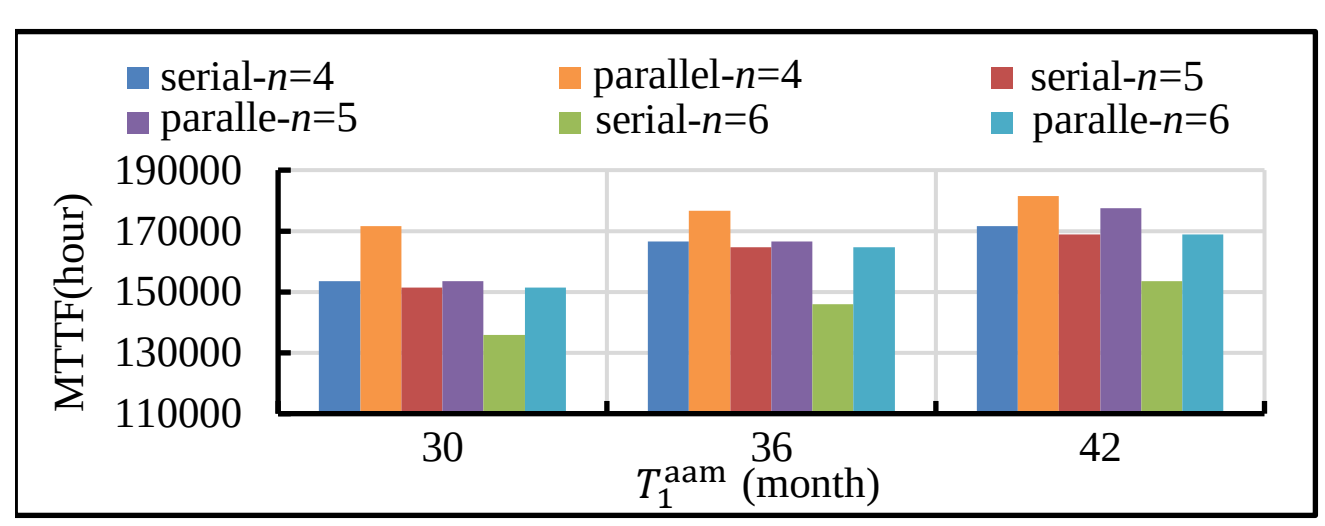

Fig. 12 The **serial and parallel MSFC MTTF** over the number of MSFs

## 4.5 Effect of Rejuvenation-triggered Intervals (RTIs) on

### MSFC Dependability

In this section, we set different RTIs to conduct numerical experiments of analyzing the effect of RTIs on MSFC dependability. TABLE 6 and TABLE 7 show the numerical results of the steady-state availability and MTTF of serial MSFC under different RTIs, respectively. The steady-state availability and MTTF of parallel MSFC are given in Section C of the supplementary file. We can observe that:

TABLE 6
SERIAL MSFC STEADY-STATE AVAILABILITY OVER RTIS

| RTI combination (hour) | Availability | RTI combination (hour) | Availability |
|---|---|---|---|
| $\omega_1^s=0,\omega_1^v=0,\omega_1^m=0$ | 0.99999797812165 | $\omega_1^s=8,\omega_1^v=0,\omega_1^m=0$ | 0.999997978165027 |
| $\omega_1^s=0,\omega_1^v=10,\omega_1^m=20$ | 0.999997978164375 | $\omega_1^s=8,\omega_1^v=10,\omega_1^m=20$ | 0.999997978225949 |
| $\omega_1^s=0,\omega_1^v=20,\omega_1^m=40$ | 0.99999797816712 | $\omega_1^s=8,\omega_1^v=20,\omega_1^m=40$ | 0.999997978230381 |
| $\omega_1^s=0,\omega_1^v=30,\omega_1^m=60$ | 0.99999797815136 | $\omega_1^s=8,\omega_1^v=30,\omega_1^m=60$ | 0.999997978204686 |
| $\omega_1^s=4,\omega_1^v=0,\omega_1^m=0$ | 0.999997978161474 | $\omega_1^s=12,\omega_1^v=0,\omega_1^m=0$ | 0.99999797814469 |
| $\omega_1^s=4,\omega_1^v=10,\omega_1^m=20$ | 0.99999797822411 | $\omega_1^s=12,\omega_1^v=10,\omega_1^m=20$ | 0.999997978194574 |
| $\omega_1^s=4,\omega_1^v=20,\omega_1^m=40$ | 0.999997978224605 | $\omega_1^s=12,\omega_1^v=20,\omega_1^m=40$ | 0.999997978197657 |
| $\omega_1^s=4,\omega_1^v=30,\omega_1^m=60$ | 0.999997978197363 | $\omega_1^s=12,\omega_1^v=30,\omega_1^m=60$ | 0.999997978165235 |

TABLE 7
SERIAL MSFC MTTF OVER RTIS

| RTI combination (hour) | MTTF (hour) | RTI combination (hour) | MTTF (hour) |
|---|---|---|---|
| $\omega_1^s=0,\omega_1^v=0,\omega_1^m=0$ | 167191.13 | $\omega_1^s=4,\omega_1^v=0,\omega_1^m=0$ | 166710.63 |
| $\omega_1^s=0,\omega_1^v=1,\omega_1^m=2$ | 167002.65 | $\omega_1^s=4,\omega_1^v=1,\omega_1^m=2$ | 166523.31 |
| $\omega_1^s=0,\omega_1^v=2,\omega_1^m=4$ | 166814.35 | $\omega_1^s=4,\omega_1^v=2,\omega_1^m=4$ | 166336.16 |
| $\omega_1^s=0,\omega_1^v=3,\omega_1^m=6$ | 166626.22 | $\omega_1^s=4,\omega_1^v=3,\omega_1^m=6$ | 166149.17 |
| $\omega_1^s=2,\omega_1^v=0,\omega_1^m=0$ | 166952.45 | $\omega_1^s=6,\omega_1^v=0,\omega_1^m=0$ | 166465.79 |
| $\omega_1^s=2,\omega_1^v=1,\omega_1^m=2$ | 166764.54 | $\omega_1^s=6,\omega_1^v=1,\omega_1^m=2$ | 166279.06 |
| $\omega_1^s=2,\omega_1^v=2,\omega_1^m=4$ | 166576.82 | $\omega_1^s=6,\omega_1^v=2,\omega_1^m=4$ | 166092.49 |
| $\omega_1^s=2,\omega_1^v=3,\omega_1^m=6$ | 166389.25 | $\omega_1^s=6,\omega_1^v=3,\omega_1^m=6$ | 165906.09 |

- The serial MSFC can achieve the maximum steady-state availability of 0.999997978240037 at $(\omega_1^s,\omega_1^v,\omega_1^m)=(6,15,30)$. As the RTIs increase, the steady-state availability of serial MSFC first increases and then decreases. Because the available time of serial MSFC system increases with the increase of RTIs when RTIs are less than the optimal point, and the probability of serial MSFC system entering the unavailable states increases with the increase of RTIs when RTIs are greater than the optimal point. At the same time, we can also observe that the MTTF of serial MSFC reaches the maximum value of 167191.136707818 at $(\omega_1^s,\omega_1^v,\omega_1^m)=(0,0,0)$ and decreases with the increase of RTIs.
- As the RTIs of parallel components increase, the steady-state availability of parallel MSFC increases, while as the RTIs of the serial components increase, the steady-state availability of parallel MSFC first increases and then decreases. The MTTF of parallel MSFC decreases with the increase of RTIs and achieves the maximum value of 172180.601132625 at $(\omega_2^s,\omega_2^v,\omega_1^m)=(0,0,0)$. As the RTIs of parallel component increase, the holding time of parallel MSFC system being available increases. Therefore, the steady-state availability of parallel MSFC system increases with the increase in the RTIs of parallel components. However, the probability of this component failure and other components aging before rejuvenation increases, leading to a decrease in the MTTF of parallel MSFC as the RTIs of parallel component increase.

## 4.6 Comparison of Our Model and the Model Without Considering Backup Components' Behaviors

In this section, we explore a scenario where backup components are not affected by software aging and failure. In order to capture the system behaviors in this scenario, we simplify the model presented in Section 3 by excluding the behaviors of backup components. Subsequently, we conduct numerical experiments to compare the proposed model with this simplified model. The experimental results are shown in Fig. 13 and Fig. 14. We can observe that the steady-state availability and MTTF under the model without considering backup component behaviors are greater than those under the model with considering backup component behaviors. For example, when $T_1^R$ (the fix time of the 1st Primary Metaverse host and its Backup Metaverse host) is 0.225 hours, the steady-state availability of serial MSFC is 0.999997946150803, and the steady-state availability of parallel MSFC is 0.999998969170332 under the model with considering backup component behaviors. Under the model without considering the behaviors of backup components, the steady-state availability of serial MSFC is 0.999998940811332, and the steady-state availability of parallel MSFC is 0.999999410993508. Therefore, whether or not to consider the behaviors of backup components has a significant impact on the outcomes.

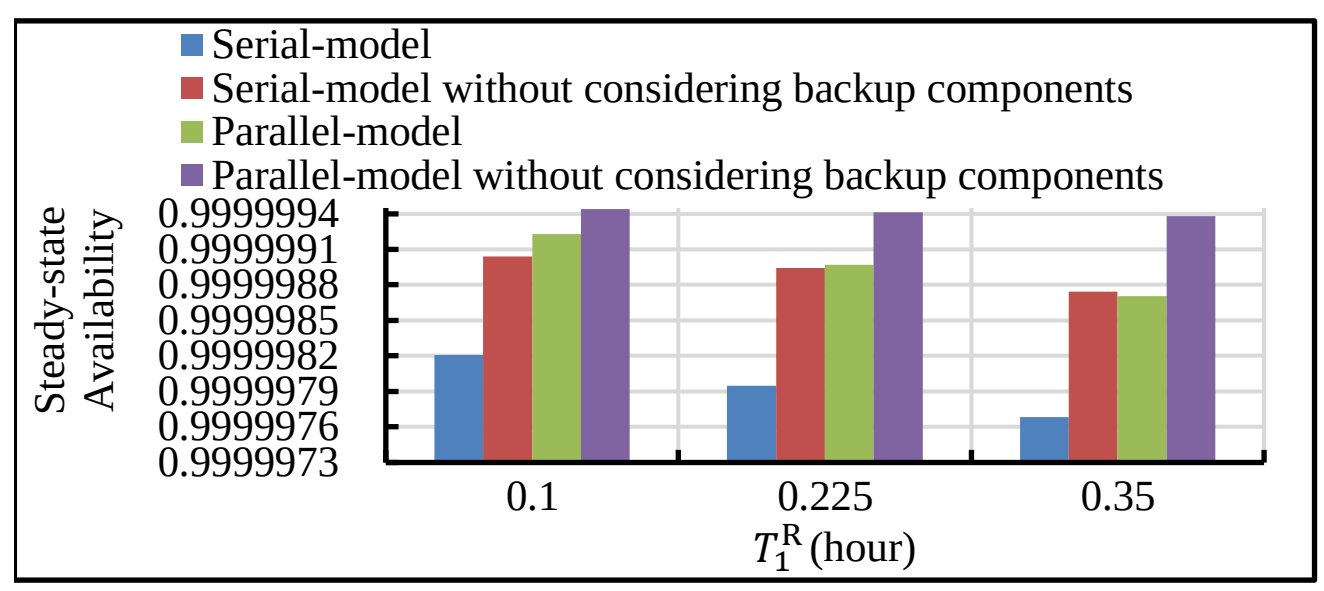

Fig. 13 Comparison of our model with the model without considering backup components in terms of MSFC steady-state availability

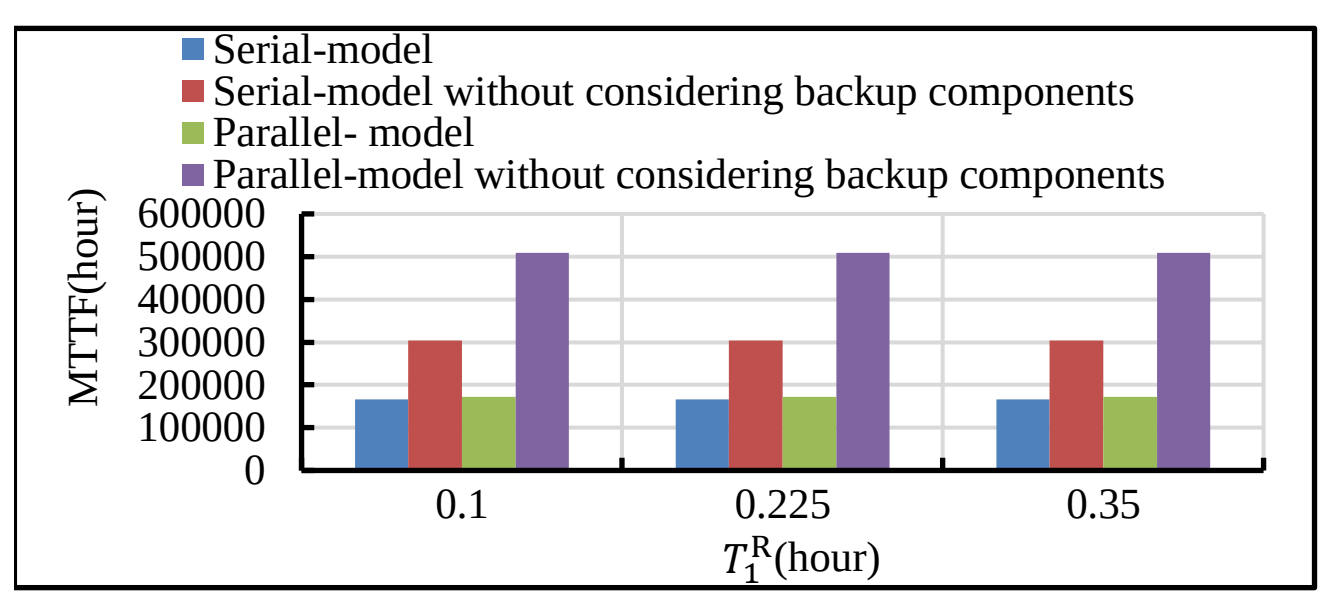

Fig. 14 Comparison of our model with the model without considering backup components in terms of MSFC MTTF

## 4.7 Effect of Cumulative Distribution Function (CDF)

### Types on MSFC Dependability

In this section, we set the $T_1^R$ (the fix time of the 1st Primary Metaverse host and its Backup Metaverse host) to vary from 0.1 hours to 0.35 hours and perform the numerical experiment of analyzing the effect of CDF types on MSFC dependability. The steady-state availability and MTTF of **serial MSFC and parallel MSFC** under different CDF types are shown in Fig. 15 and Fig. 16, respectively. In these figures, 'S-F_HYPO_R_EXP' and 'P-F_HYPO_R_EXP' denote that all failure times follow the hypoexponential distribution and all recovery times follow the exponential distribution for serial and parallel MSFC, respectively. 'S-F_HYPO_R_Deter' and 'P-F_HYPO_R_Deter' denote that all failure times follow the hypoexponential distribution and all recovery times follow the deterministic distribution for serial and parallel MSFC, respectively. 'S-F_EXP_R_EXP' and 'P-F_EXP_R_EXP' denote that all failure times and recovery times follow the exponential distribution for serial and parallel MSFC, respectively. 'S-F_EXP_R_Deter' and 'P-F_EXP_R_Deter' denote that all failure times follow the exponential distribution and all recovery times follow the deterministic distribution for serial MSFC and parallel MSFC, respectively. We can observe that CDF type of failure time is an important factor for improving MSFC dependability.

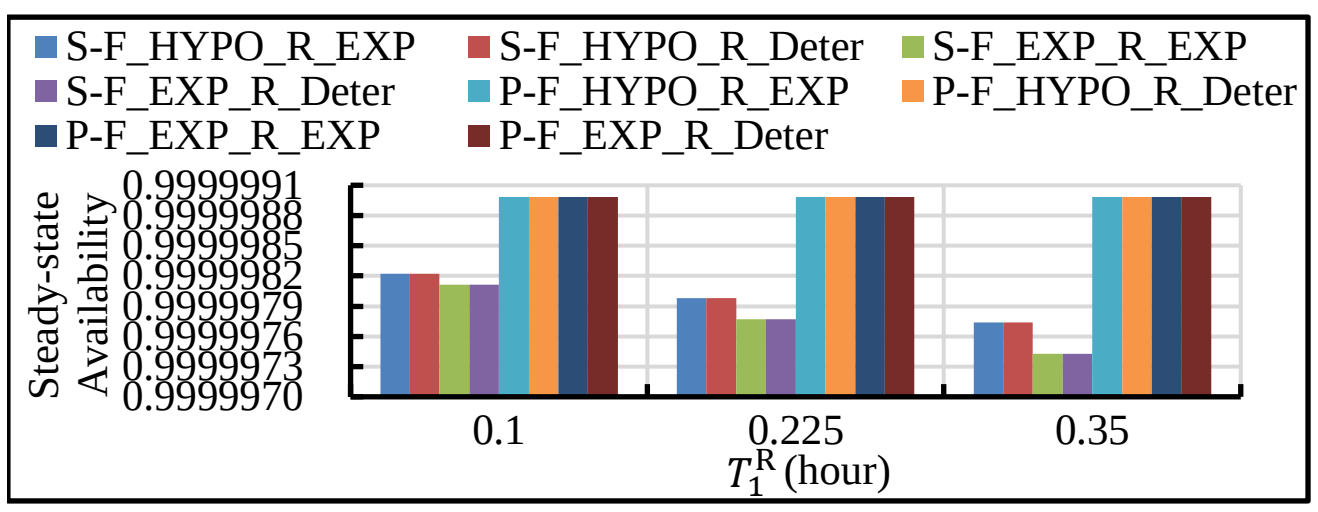

Fig. 15 MSFC steady-state availability over $T_1^R$ and CDF types of failure time and recovery time

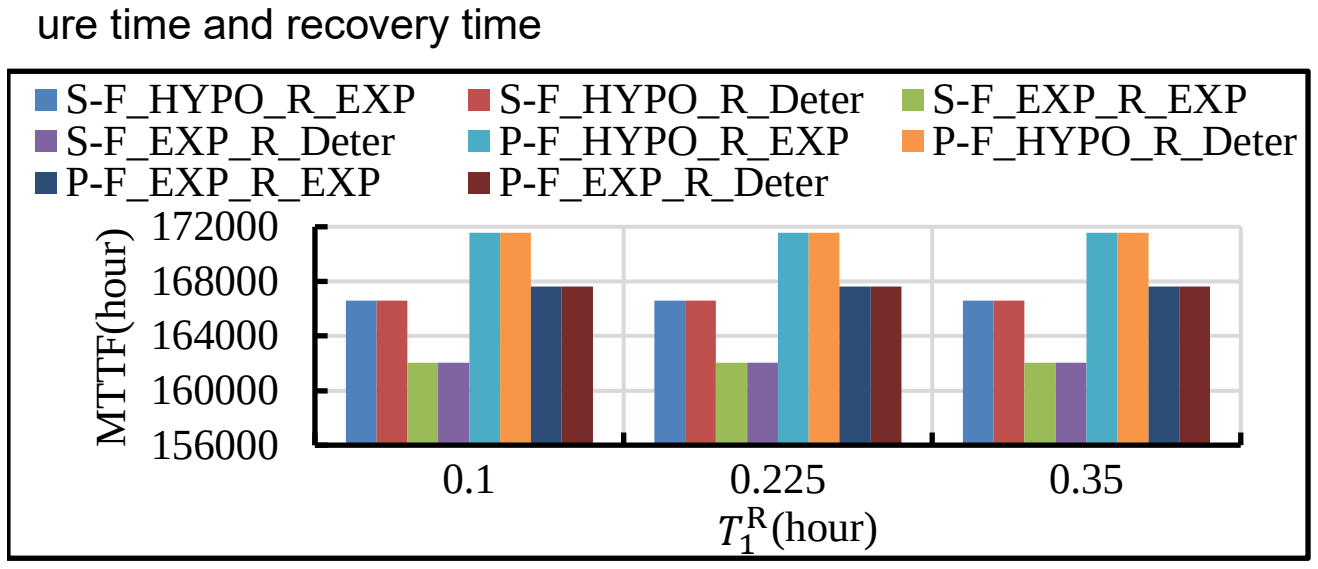

Fig. 16 MTTF of MSFC systems over $T_1^R$ and CDF types of failure time and recovery time

## 5 Conclusion

In this paper, we develop a hierarchical model of an MSFC system consisting of $n$ MSFs. There are $n$ SMP sub-models and a RBD sub-model in the model. Each SMP sub-model desribes the behaviors of components in a Primary Metaverse host and its Backup Metaverse host, and the RBD sub-model describes the composition of $n$ SMP sub-models. Then, we derive the closed-formed formulas of calculating the steady-state availability and reliability of an MSFC comprising arbitrary number of MSFs. Finally, we evaluate the effect of system parameter, backup components' behaviors and CDF types on MSFC dependability, providing guidance for dependability optimization.

This paper assumes that an active component has a corresponding backup component. Nevertheless, it is possible that an active component has multiple corresponding backup components. In future work, we will explore the influence of the number of backup components on MSFC dependability.

## References

[1] "Introducing Meta: A social technology company," [Online]. Available: https://about.fb.com/news/2021/10/facebook-company-is-now-meta/.

[2] "Mesh for Microsoft Teams aims to make collaboration in the 'Metaverse' personal and fun," [Online]. Available: https://news.microsoft.com/source/features/innovation/mesh-for-microsoft-teams/.

[3] Yi-Jing Liu, Hongyang Du, Dusit Niyato, Gang Feng, Jiawen Kang, Zehui Xiong: Slicing4Meta: An Intelligent Integration Architecture with Multi-Dimensional Network Resources for Metaverse-as-a-Service in Web 3.0. IEEE Commun. Mag. 61(8): 20-26 (2023).

[4] Hang Wang, Huansheng Ning, Yujia Lin, Wenxi Wang, Sahraoui Dhelim, Fadi Farha, Jianguo Ding, Mahmoud Daneshmand: A Survey on the Metaverse: The State-of-the-Art, Technologies, Applications, and Challenges. IEEE Internet Things J. 10(16): 14671-14688 (2023).

[5] Nam H. Chu, Dinh Thai Hoang, Diep N. Nguyen, Khoa T. Phan, Eryk Dutkiewicz: MetaSlicing: A Novel Resource Allocation Framework for Metaverse. IEEE Transactions on Mobile Computing ( Early access, 2023).

[6] https://hanu.com/why-every-successful-metaverse-initiative-needs-a-great-right-cloud-provider/

[7] Kengo Watanabe, Fumio Machida, Ermeson Carneiro de Andrade, Roberto Pietrantuono, Domenico Cotroneo: Software Aging in a Real-Time Object Detection System on an Edge Server. SAC 2023: 671-678.

[8] Yaru Li, Lin Li, Jing Bai, Xiaolin Chang, Yingying Yao, Peide Liu: Availability and Reliability of Service Function Chain: A Quantitative Evaluation View. Int. J. Comput. Intell. Syst. 16(1): 52 (2023).

[9] Jiao Zhang, Zenan Wang, Chunyi Peng, Linquan Zhang, Tao Huang, Yunjie Liu: RABA: Resource-Aware Backup Allocation For A Chain of Virtual Network Fnctions. INFOCOM 2019: 1918-1926.

[10] Meng Wang, Bo Cheng, Shangguang Wang, Junliang Chen: Availability- and Traffic-Aware Placement of Parallelized SFC in Data Center Networks. IEEE Trans. Netw. Serv. Manag. 18(1): 182-194 (2021).

[11] Chen Wang, Qin Hu, Dongxiao Yu, Xiuzhen Cheng: Online Learning for Failure-aware Edge Backup of Service Function Chains with the Minimum Latency. IEEE Transactions on Networking (2023).

[12] Lanlan Rui, Xushan Chen, Zhipeng Gao, Wenjing Li, Xuesong Qiu, Luoming Meng: Petri Net-Based Reliability Assessment and Migration Optimization Strategy of SFC. IEEE Trans. Netw. Serv. Manag. 18(1): 167-181 (2021).

[13] Luigi De Simone , Mario Di Mauro , Roberto Natella , Fabio Postiglione : A Latency-Driven Availability Assessment for Multi-Tenant Service Chains. IEEE Trans. Serv. Comput. 16(2): 815-829 (2023).

[14] Besmir Tola, Yuming Jiang, Bjarne E. Helvik: Model-Driven Availability Assessment of the NFV-MANO With Software Rejuvenation. IEEE Trans. Netw. Serv. Manag. 18(3): 2460-2477 (2021).

[15] Jing Bai, Xiaolin Chang, Fumio Machida, Lili Jiang, Zhen Han, Kishor S. Trivedi: Impact of Service Function Aging on the Dependability for MEC Service Function Chain. IEEE Trans. Dependable Secur. Comput. 20(4): 2811-2824 (2023).

[16] Jing Bai, Xiaolin Chang, Kishor S. Trivedi, Zhen Han: Resilience-Driven Quantitative Analysis of Vehicle Platooning Service. IEEE Trans. Veh. Technol. 70(6): 5378-5389 (2021).

[17] Jing Bai, Xiaolin Chang, Fumio Machida, Zhen Han, Yang Xu, Kishor S. Trivedi: Quantitative understanding serial-parallel hybrid sfc services:

a dependability perspective. Peer-to-Peer Netw. Appl. 15(4): 1923-1938 (2022).

[18] Jing Bai, Xiaolin Chang, Ricardo J. Rodríguez, Kishor S. Trivedi, Shupan Li: Towards UAV-Based MEC Service Chain Resilience Evaluation: A Quantitative Modeling Approach. IEEE Trans. Veh. Technol. 72(4): 5181-5194 (2023).

[19] Jing Bai, Xiaolin Chang, Fumio Machida, Kishor S. Trivedi, Yaru Li: Model-Driven Dependability Assessment of Microservice Chains in MEC-Enabled IoT. IEEE Trans. Serv. Comput. 16(4): 2769-2785 (2023).

[20] Jing Bai, Yaru Li, Xiaolin Chang, Fumio Machida, Kishor S. Trivedi: Understanding NFV-Enabled Vehicle Platooning Application: A Dependability View. IEEE Trans. Cloud Comput. (2023).

[21] Mario Di Mauro, Giovanni Galatro, Maurizio Longo, Fabio Postiglione, Marco Tambasco: Comparative Performability Assessment of SFCs: The Case of Containerized IP Multimedia Subsystem. IEEE Trans. Netw. Serv. Manag. 18(1): 258-272 (2021).

[22] Thilina Pathirana, Gianfranco Nencioni: Availability Model of a 5G-MEC System. CoRR abs/2304.09992 (2023).

[23] Jinxia Wang, Stanislav Makowski, Alan Cieślik, Haibin Lv, Zhihan Lv: Fake News in Virtual Community, Virtual Society, and Metaverse: A Survey. IEEE Trans. Comput. Social Syst. (2023).

[24] Thien Huynh-The, Thippa Reddy Gadekallu, Weizheng Wang, Gokul Yenduri, Pasika Ranaweera, Quoc-Viet Pham, Daniel Benevides da Costa, Madhusanka Liyanage: Blockchain for the metaverse: A Review. Future Gener. Comput. Syst. 143: 401-419 (2023).

[25] Ersin Dincelli, Alper Yayla: Immersive virtual reality in the age of the Metaverse: A hybrid-narrative review based on the technology affordance perspective. J. Strateg. Inf. Syst. 31(2): 101717 (2022).

[26] Zhihan Lv, Shuxuan Xie, Yuxi Li, M. Shamim Hossain, Abdulmotaleb El-Saddik: Building the Metaverse by Digital Twins at All Scales, State, Relation. Virtual Real. Intell. Hardw. 4(6): 459-470 (2022).

[27] Seok W H: Analysis of metaverse business model and ecosystem[J]. Electronics and Telecommunications Trends, 36(4): 81-91(2021).

[28] Thien Huynh-The, Quoc-Viet Pham, Xuan-Qui Pham, Thanh Thi Nguyen, Zhu Han, Dong-Seong Kim: Artificial intelligence for the metaverse: A survey. Eng. Appl. Artif. Intell. 117(Part): 105581 (2023).

[29] Kai Li, Yingping Cui, Weicai Li, Tiejun Lv, Xin Yuan, Shenghong Li, Wei Ni, Meryem Simsek, Falko Dressler: When Internet of Things Meets Metaverse: Convergence of Physical and Cyber Worlds. IEEE Internet Things J. 10(5): 4148-4173 (2023).

[30] MapleSoft, "Maple," [Online]. Available: http://www.maplesoft.com/products/maple.